\documentclass[traditabstract]{aa}

\usepackage{graphicx}
\usepackage{txfonts}
\usepackage{natbib}
\usepackage{enumerate}
\usepackage{bm}
\usepackage{hyperref}
\hypersetup{
  colorlinks=true,
  linkcolor=blue,
  citecolor=blue,
  urlcolor=blue,
}

\bibpunct{(}{)}{;}{a}{}{,} 

\newcommand{\sbse}{SBS\,0335$-$052\,E}
\newcommand{\sbsw}{SBS\,0335$-$052\,W}
\newcommand{\sbs}{SBS\,0335$-$052}
\newcommand{\izw}{I\,Zw\,18}

\newcommand{\dgr}{$\mathcal{D}$}
\newcommand{\coone}{$^{12}$CO(1--0)}
\newcommand{\cotwo}{$^{12}$CO(2--1)}

\newcommand{\aco}{$\alpha_{\rm CO}$}

\newcommand{\kms}{km\,s$^{-1}$}
\newcommand{\kkms}{K\,km\,s$^{-1}$}
\newcommand{\htwo}{H$_2$}
\newcommand{\av}{A$_V$}
\newcommand{\spit}{{\it Spitzer}}
\newcommand{\hers}{{\it Herschel}}
\newcommand{\hst}{{\sl HST}}
\newcommand{\iso}{{\it ISO}}
\newcommand{\iras}{{\it IRAS}}

\newcommand{\dusty}{{\it DUSTY}}
\newcommand{\logoh}{12$+$log(O/H)}

\newcommand{\sigmasfr}{$\Sigma_{\rm SFR}$}
\newcommand{\sigmadust}{$\Sigma_{\rm dust}$}

\newcommand{\sigmahtwo}{$\Sigma_{\rm H2}$}
\newcommand{\mstar}{$M_{\rm star}$}
\newcommand{\mdust}{$M_{\rm dust}$}
\newcommand{\hi}{\rm H{\sc i}}

\newcommand{\pa}{${\rm Pa}\alpha$}

\newcommand{\bra}{${\rm Br}\alpha$}
\newcommand{\brg}{${\rm Br}\gamma$}
\newcommand{\ha}{${\rm H}\alpha$}

\newcommand{\micron}{$\mu$m}
\newcommand{\tin}{T$_{\rm in}$}

\newcommand{\taud}{$\tau_{\rm dust}$}
\newcommand{\tauvd}{$\tau_{\rm V}^{\rm dust}$}

\def\tex {\ifmmode{{T}_{\rm ex}}\else{$T_{\rm ex}$}\fi}
\def\tmb {\ifmmode{{T}_{\rm mb}}\else{$T_{\rm mb}$}\fi}
\def\ci  {\ifmmode{{\rm C}{\rm \small I}}\else{C\ts {\scriptsize I}}\fi}
\def\hi  {\ifmmode{{\rm H}{\rm \small I}}\else{H\ts {\scriptsize I}}\fi}
\def\hh  {\ifmmode{{\rm H}_2}\else{H$_2$}\fi}
\def\kms    {\ifmmode{{\rm \ts km\ts s}^{-1}}\else{\ts km\ts s$^{-1}$}\fi}
\def\msun   {\ifmmode{{\rm M}_{\odot}}\else{M$_{\odot}$}\fi}
\def\msunpc   {\ifmmode{{\rm M}_{\odot}\,{\rm pc}^{-2}}\else{M$_{\odot}$\,pc$^{-2}$}\fi}
\def\msunyr   {\ifmmode{{\rm M}_{\odot}\,{\rm yr}^{-1}}\else{M$_{\odot}$\,yr$^{-1}$}\fi}
\def\lsun   {\ifmmode{{\rm L}_{\odot}}\else{L$_{\odot}$}\fi}
\def\zsun   {\ifmmode{{\rm Z}_{\odot}}\else{Z$_{\odot}$}\fi}

\begin{document}

\title{ALMA observations of cool dust in a low-metallicity starburst, \sbs\
\thanks{Based on observations carried out with ALMA in Cycle 0;
the Joint ALMA Observatory is operated by ESO, AUI/NRAO and NAOJ.}
}

\author{L. K. Hunt \inst{1}
\and
L. Testi \inst{1,2}
\and
V. Casasola \inst{3}
\and
S. Garc\'{\i}a-Burillo \inst{4}
\and
F. Combes \inst{5}
\and
R. Nikutta \inst{6}
\and
P. Caselli \inst{7}
\and
C. Henkel \inst{8,9}
\and
R. Maiolino \inst{10}
\and
K. M. Menten \inst{8}
\and
M. Sauvage \inst{11}
\and
A. Weiss \inst{8}
}

\offprints{L. K. Hunt}
\institute{INAF - Osservatorio Astrofisico di Arcetri, Largo E. Fermi, 5, 50125, Firenze, Italy
\email{hunt@arcetri.astro.it}
\and
ESO, Karl Schwarzschild str. 2, 85748 Garching bei M\"unchen, Germany
 \and
INAF -- Istituto di Radioastronomia \& Italian ALMA Regional Centre, via Gobetti 101, 40129, Bologna, Italy
 \and
Observatorio Astron\'omico Nacional (OAN)-Observatorio de Madrid,
Alfonso XII, 3, 28014-Madrid, Spain
 \and
Observatoire de Paris, LERMA (CNRS:UMR8112), 61 Av. de l'Observatoire, F-75014, Paris, France
 \and
Universidad Andr{\' e}s Bello, Deparamento de Ciencias F{\' i}sicas, Rep{\' u}blica
252, Santiago, Chile
 \and
School of Physics and Astronomy, University of Leeds, Leeds LS2 9JT, UK 
 \and
Max-Planck-Institut f\"ur Radioastronomie, Auf dem H\"ugel 69, 53121, Bonn, Germany
 \and
Astronomy Department, King Abdulaziz University, P.O. Box 80203, Jeddah, Saudia Arabia
 \and
Cavendish Laboratory, University of Cambridge, 19 J.J. Thomson Avenue, Cambridge CB3 0HE, UK
 \and
CEA, Laboratoire AIM, Irfu/SAp, Orme des Merisiers, 91191 Gif-sur-Yvette, France
}

   \date{Received  2013/ Accepted  2013}

   \titlerunning{Cool dust in SBS\,0335$-$052}
   \authorrunning{Hunt et al.}

\abstract{We present Atacama Large Millimeter/submillimeter Array (ALMA) Cycle 0 Band 7 observations of an extremely
metal-poor dwarf starburst galaxy in the Local Universe, \sbs\ (\logoh\,$\sim$\,7.2).
With these observations, dust is detected at 870\,\micron\ (ALMA Band 7),
but 85\% of the flux in this band is due to free-free emission from the starburst.
We have compiled a spectral energy distribution (SED) of \sbs\ that spans
almost 6 orders of magnitude in wavelength and fit it with
a spherical dust shell heated by a single-age stellar population; the best-fit
model gives a dust mass of $(3.8\,\pm\,0.6)\times10^4$\,\msun.
We have also constructed a SED including \hers\ archival data 
for \izw, another low-metallicity dwarf starburst (\logoh\,$\sim$\,7.17),
and fit it with a similar model to
obtain a dust mass of $(3.4\,\pm\,1.0)\times10^2$\,\msun.
It appears that for such low-metallicity dwarf galaxies, the ratio of
their stellar mass to their dust mass is within the range of values found for spirals and
other star-forming galaxies.
However, compared with their atomic gas mass, the dust mass of \sbs\ far exceeds the prediction
of a linear trend of dust-to-gas mass ratio with metallicity, while \izw\ falls
far below.
We use gas scaling relations to assess a putative missing gas component in both
galaxies and find that the missing, possibly molecular, gas in \sbs\ is a factor
of 6 times higher than the value inferred from the observed \hi\ column density; 
in \izw\ the missing component is much smaller.
Finally, we constrain the \htwo\ surface density conversion factor \aco\ with our upper limit
for CO\,$J=3-2$ line in \sbs, and find that this is consistent with a linear or even
super-linear trend of increasing \aco\ with decreasing metallicity.
Ultimately, despite their similarly low metallicity, 
the differences in gas and dust column densities in \sbs\ and \izw\
suggest that metal abundance does not uniquely define star-formation processes. 
At some level, self-shielding and the survival of molecules may depend
just as much on gas and dust column density as on metallicity.
The effects of low metallicity may at least be partially compensated for
by large column densities in the interstellar medium.
}
\maketitle


\section{Introduction}
\label{sec:intro}

First \iras\ and \iso, then SCUBA, {\it COBE}, and most recently
\spit\ and \hers\ have convincingly shown that most of
the star formation in the universe is enshrouded in dust.
Large populations of infrared-luminous galaxies are major contributors
to the far-infrared and sub-millimeter backgrounds, and
are probably responsible for most of
the star-formation activity at high redshifts
\citep[e.g.,][]{chary01,dole06,gruppioni10}.
Indeed, half the energy and most of the photons in 
the universe come from the infrared spectral region
\citep[e.g.,][]{hauser01,franceschini08}.

That dust is so prominent in the high-redshift universe
may appear surprising, since it has been assumed that
dust would be absent in primordial, 
metal-poor environments.
However, the presence of large amounts of dust (and molecular gas) in $z\sim6$ quasars
\citep[e.g.,][]{wang08} 
has been attributed to 
supernovae and Asymptotic Giant Branch stars which evolve in a metal-free
interstellar medium (ISM) \citep[e.g.,][]{todini01,schneider04,nozawa07,valiante09}.
Dust at such high redshifts
suggests that early star formation episodes must be very intense
and relatively brief.
However, exactly how these massive starbursts occur and evolve
is not yet clear.
The short interval in which star formation and the ensuing chemical enrichment
and dust formation convert a dust-free metal-free environment to a dusty
metal-rich one by redshift $z\sim$6 is as yet unobserved, and studies of 
such transitions remain a major observational challenge. 

The Local Universe is home to star-forming dwarf galaxies that are much more 
metal poor than galaxies observed thus far at high redshift. 
Only a handful of blue compact dwarf (BCD) galaxies have been discovered
so far with \logoh$\sim$7.2 \citep[e.g.,][]{izotov07}. 
Even at these extremely low metallicities,
BCD spectral energy distributions (SEDs) can be dominated by 
infrared dust emission resulting from reprocessing of UV radiation
from young massive stars
\citep[e.g., \sbs\ and \izw:][]{houck04a,wu07}. 
Because such galaxies are chemically unevolved, they 
can provide a window on primordial galaxy formation and evolution.
This is, in some sense, a ``local'' approach to a cosmological problem.
If we can study the properties of a metal-poor ISM and its constituents
locally, we may be able to better understand the high-redshift transition from
metal-free Population\,III stars to the chemically evolved
massive galaxies typical of the current epoch.

In this paper, we present Atacama Large Millimeter/submillimeter Array (ALMA) 
Cycle 0 Band 7 (346.21\,GHz) observations of an extremely
low-metallicity starburst in the Local Universe, \sbs.
We also compare it with a nearby similarly metal-poor dwarf galaxy, \izw.

\subsection{SBS0335$-$052}
\label{sec:sbs}

\sbs\ (more properly, the eastern component, \sbse) is embedded together with a western component,
\sbsw, in a giant \hi\ cloud roughly 4.2\,arcmin across,
or 66\,kpc at \sbs's distance \citep[][]{pustilnik01,ekta09}.
\sbsw, although physically related to \sbse\ because of their common \hi\ envelope,
is 22\,kpc (1.4\,arcmin) distant from \sbse.
Both galaxies are very metal poor:
\logoh$\sim$7.2 for \sbse\ and 7.12 for \sbsw,
similar to the ``prototype'' BCD, \izw\ with \logoh\,=\,7.17 
\citep{izotov99,izotov05}.
\sbse\ hosts six Super Star Clusters (SSCs), 
distributed (end-to-end) over roughly 2\farcs6 ($\sim$700\,pc), 
with most of the star formation activity taking place in the two brightest ones 
(SSC\,1, 2) to the southeast \citep{hunt01,dale01}. 
There are more than 10000 O stars within
these two clusters, unresolved at {\it HST} resolution \citep{thuan97b,reines08}.
The stars in \sbs\ are mostly confined to the six SSCs, with a total mass of 
$\sim 6\times10^6$\,\msun\ \citep{reines08}.
However, there is a faint underlying stellar component with
a mass comparable to the sum of the SSCs; \citet{papaderos98} estimate a total mass
including the diffuse populations of $3.1\times10^6$\,\msun, roughly
half of that given by Reines et al. for the clusters alone.
Hence, in this paper we adopt the more massive estimate of \citet{reines08}.

The total amount of \hi, $\sim 4.3\times10^8$\,\msun\ in \sbse\ makes it extremely
gas rich, even without a, as yet undetected, \htwo\ component \citep{ekta09}.
The global radio continuum shows a non-thermal slope \citep{hunt04}, but with
significant free-free absorption (i.e., by dense gas) which is even more 
prominent at high spatial resolution. 
Compact rising-spectrum (optically thick) thermal radio sources are found in the
vicinity of SSCs 1 and 2 \citep{johnson09}, together with \pa\ \citep{reines08}.
Even the optical spectrum implies elevated ionized gas densities $\ga$500\,cm$^{-3}$ \citep{izotov99}.
In the relation between star-formation rate (SFR) surface density \sigmasfr,
and gas surface density (the Schmidt-Kennicutt law), \sbse\
is the equivalent of a massive starburst, 
with SFR\,=\,1.2\,\msun\,yr$^{-1}$ occurring
in a region of $<$500\,pc in diameter. 
Because
of their physical and apparent separation, hereafter we will refer to the eastern component,
\sbse, as \sbs\ without the explicit ``E'' suffix. 
Table \ref{tab:basicsbs} shows the basic parameters for \sbs.
At the assumed distance of $\sim$54.1\,Mpc, 1\arcsec\,=\,262\,pc.

\subsection{Another metal-poor compact dwarf galaxy, I\,Zw\,18}
\label{sec:izwintro}

In order to place \sbs\ in context, in this paper we compare its dust and gas
properties with another very low-metallicity star-forming dwarf galaxy,
\izw.

\izw\ is closer than \sbs, at a distance of $\sim$18\,Mpc, and
like \sbs, is embedded together with a faint companion, the ``C'' component,
in a massive \hi\ envelope,
although smaller in extent than that of \sbs\ \citep[$\sim$16\,kpc,][]{lelli12}.
Also like \sbs, \izw\ is extremely gas rich, with $\sim 10^8$\,\msun\ of atomic gas. 
Star formation in \izw\ takes place in two massive star clusters,
which together host $\sim$1900 massive stars.
These clusters are more extended than those in \sbs\footnote{Again,
as throughout the paper, we are referring to \sbs E even though the suffix
is omitted.} ($\sim$150$-$200\,pc in diameter),
similar to concentrations of normal OB associations in the Milky Way and the
Large Magellanic Cloud \citep{hunter95}\footnote{The conclusions of \citet{hunter95}
are still valid even with the updated distance of \izw.}.
The mass of stars in \izw, $\sim 1.3\times10^6$\,\msun\ \citep{fumagalli10}, is 
about 4 times lower than in \sbs, and
\izw's SFR, estimated from radio free-free emission \citep{hunt05b}
is 0.17\,\msunyr, about 5 times lower than in \sbs.
The SFR surface density \sigmasfr\ is more than 30 times lower than in \sbs.
Unlike \sbs, there is no evidence for self-absorption in the radio spectrum
of \izw\ \citep{hunt05b,cannon05}, implying the absence of dense gas.

Both \sbs\ and \izw\ are thought to be interacting
systems, embedded within their common \hi\ envelope \citep[e.g.,][]{ekta08}.
This could foster star formation in massive central star clusters
as modeled by \citet{bekki08}, perhaps causing age gradients 
in the stellar populations as observed \citep[e.g.,][]{thuan97b,reines08,contreras11}. 

\subsection{Constraining cool dust in SBS0335$-$052}
\label{sec:sbsintro}

Although Balmer lines suggest that the extinction \av\ in \sbs\ is $\sim$0.5\,mag
\citep{thuan97b,vanzi00}, the near-infrared Brackett line ratios (\brg/\bra) 
imply \av$\,\sim$\,12\,mag \citep{hunt01}.
The form of the infrared (IR) SED also indicates high values \citep[\av$\sim$15\,mag,][]{houck04a}, 
but detailed studies of the optical SEDs of the individual SSCs show that the dust 
must be highly clumped \citep{reines08}.
Up to now, the total dust mass was highly uncertain \citep{plante02,hunt05a}, mainly because
the SED peaks at $\sim$30\,\micron\ \citep{houck04a}, and there were no 
data available to constrain the mass in the cooler dust expected to 
dominate at longer wavelengths.
Cool dust constitutes the largest contribution to the dust mass budget,
so observations at wavelengths well longward of the peak of the 
SED are crucial for reliable estimates of dust mass.

In this paper, we add this critical information via observations of the 870\,\micron\
continuum emission in this low-metallicity BCD.
In Sect. \ref{sec:data}, we present the data we have acquired and compiled in order
to better constrain the dust mass and gas content in \sbs.
We then estimate the free-free contribution to the SED,
and fit the corrected multi-wavelength SED in Sect. \ref{sec:sed} in order to estimate the total
dust mass. 
In Sect. \ref{sec:izw}, we fit the SED of \izw\ in the same way as for \sbs,
and derive the dust mass and compare it with the gas content of \izw\ and with \sbs.
Finally, in Sect. \ref{sec:dustgas} we discuss the implied dust-to-gas ratio in \sbs\ and \izw\
and compare them with other low-metallicity dwarfs and normal spiral galaxies.
We also constrain the CO-to-\htwo\ conversion factor and infer limits to molecular 
gas mass from gas scaling relations.
Our conclusions are given in Sect. \ref{sec:conclusions}.

%
\begin{center}
\begin{table}
      \caption[]{Basic data for \sbs} 
\label{tab:basicsbs}
\addtolength{\tabcolsep}{7pt}
{
\tiny
\begin{tabular}{llc}
\hline
Parameter  & \multicolumn{1}{c}{Value$^{\mathrm{a}}$} & Reference$^{\mathrm{b}}$ \\
\hline
$V_{\rm hel}$  & 4053 km\,s$^{-1}$           & (1) \\
Distance       & 54.1\,Mpc (1\arcsec\ = 262\,pc)   & (1) \\
\logoh         & 7.19 (toward SW) $-$ 7.34 (NE)    & (2) \\
M$_{\rm stars}$    & $5.68 \times 10^{6}$\,\msun & (3) \\
SFR  	       & 1.2\,\msunyr                  & (4) \\
$\Sigma_{\rm SFR}^{\mathrm{c}}$ & 5\,\msunyr\,kpc$^{-2}$  & This paper \\
M$_{\rm H\,I}$ & $4.3 \times 10^{8}$\,\msun  & (5) \\
$\Sigma_{\rm H\,I}$ & 56\,\msunpc  & (6) \\
L$_{\rm dust}^{\mathrm{d}}$  & $(1.65\,\pm\,0.1) \times 10^{9}$\,\lsun  & This paper \\
M$_{\rm dust}$  & $(3.8\,\pm\,0.6) \times 10^{4}$\,\msun  & This paper \\
\hline
\end{tabular}
}
\vspace{0.5\baselineskip}
\begin{description}
\item
[$^{\mathrm{a}}$]
Luminosity and mass values taken from the literature
have been scaled to the distance of $D$ = 54.1\,Mpc. 
\item
[$^{\mathrm{b}}$] (1) NASA/IPAC Extragalactic Database (NED, \\
http://nedwww.ipac.caltech.edu/) Luminosity Distance
scaled to the reference frame defined by the 3K Microwave Background Radiation;
(2) \citet{izotov99};
(3) \citet{reines08};
(4) \citet{johnson09};
(5) In the entire \sbse\ region \citep{ekta09};
(6) \citet{thuan97a}.
\item
[$^{\mathrm{c}}$]
Obtained by considering an area of 0.2\,kpc$^2$,
corresponding to $\sim 1\farcs9\times1\farcs9$\,arcsec$^2$.
This is a very conservative estimate, since the star formation
takes place in the two brightest SSCs, unresolved at an
\hst\ resolution of $\sim$20\,pc.
\item
[$^{\mathrm{d}}$]
Obtained by integrating the SED model described in Sect. \ref{sec:sed}.
\end{description}
\end{table}
\end{center}

\section{Data for SBS0335$-$052}
\label{sec:data}

We have acquired Band 7 images with ALMA
in Cycle 0 and also retrieved data from the 
\hers\ archive, and combined them with data from the literature to compile
a SED for \sbs\ that spans almost 6 orders of magnitude in wavelength.

\subsection{ALMA observations}
\label{sec:alma}

The observations were carried out in five separate 
sessions from Jul 30 through Aug 14, 2012 with the ALMA observatory as part of Early Science Cycle 0.
The average number of antennas was 24 and the array was in the Early Science extended configuration
providing baselines up to 400\,m.
\sbs\ was observed with Band 7, simultaneously 
in the continuum and in the $^{12}$CO(3-2) transition, with a 
rest frequency of 345.79599~GHz.
The central sky frequencies of the upper and lower sidebands were
$\sim 352$~GHz and $\sim 340$~GHz, respectively. The observations
were done as 6 executions of the same scheduling block.
The unresolved extragalactic source J0339$-$017 was used for complex gain calibration,
while Callisto or Uranus were used as flux calibrators. The measured fluxes 
of J0339$-$017 during the various observing sessions were consistently found
to be $\sim 0.98$~Jy and $\sim 0.95$~Jy at 340~GHz and 352~GHz, respectively.
The flux density scale accuracy is expected to be better than 5\%.

The full width at half maximum of the  ALMA antennas' primary beam at the observing frequency is $\sim17$~arcsec. 
The galaxy was observed
in dual polarization mode with 1.875~GHz total bandwidth per baseband, and
a velocity resolution of $\sim$0.9~\kms. The spectra were then smoothed
to 10~\kms\ to build channel maps.
The original plan was to use the line free part of the band to measure 
continuum emission. As no emission was detected in the CO(3-2) line, 
in the end,
the full 7.5~GHz bandwidth per polarization was used to obtain a continuum measurement centered
at the average frequency of $\sim 346$~GHz.
The data were reduced, calibrated and imaged with the CASA software \citep[v3.4,][]{mcmullin07}.
The final maps have a beam size of 0\farcs7$\times$0\farcs45 and PA of 79$^\circ$, with
a $1\sigma$ rms of 0.035\,mJy/beam in the continuum.

No CO(3-2) emission was detected.
Our ALMA measurement gives a 1$\sigma$ limit of 0.8\,mJy/beam
in 10\,\kms\ channels, which translates to 23.5\,mK.
Assuming the CO line width in \sbs\ is the same as that for
\hi\ ($\sim$50\,\kms), we obtain a 3$\sigma$ upper limit on the CO(3-2) flux
of 1.58\,\kkms.
Assuming an area of 1\,arcsec$^2$ subtended by the
source (see Sect. \ref{sec:dustgasratios}), 
and a CO(3-2)/CO(1-0) flux ratio of 0.6 (see Sect. \ref{sec:alphaco}),
this would give a 3$\sigma$ limit of the luminosity in the CO(1-0) line
$L^\prime_{\rm CO}\,=\,2\times10^{5}$\,\kkms\,pc$^{-2}$.

\begin{figure}[ht]
\centerline{
\includegraphics[angle=0,width=\linewidth]{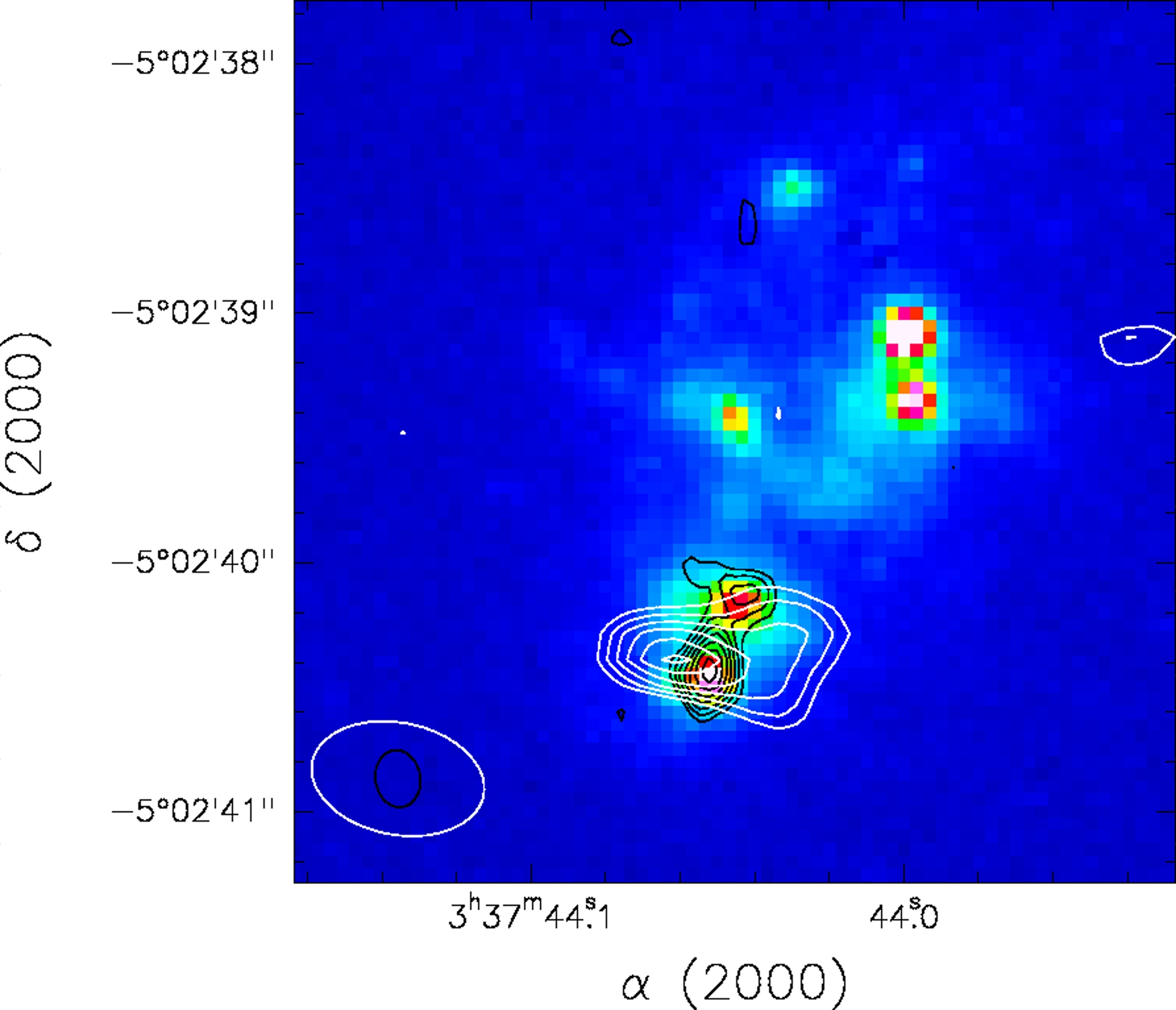}
}
\caption{ALMA 870\,\micron\ continuum map of \sbs\ shown as white contours overlaying the \hst/ACS F555M
image taken from \citet{reines08}.
Band\,7 contours range from 1 to 2$\sigma$ ($\sigma$\,=\,0.035\,mJy/beam).
Black contours (ranging from 3 to 9$\sigma$ ($\sigma$\,=\,0.012\,mJy/beam)
give the high-resolution VLA map at 3.6\,cm taken from \citet{johnson09}. 
The radio and dust emission is associated with the two brightest SSCs
(1, 2) toward the southeast.
Beam sizes of the two contoured images are shown in the lower left corner.
At the distance of \sbs, 1\arcsec\ corresponds to 262\,pc.
}
\label{fig:sbsalma}
\end{figure}

Figure \ref{fig:sbsalma} shows the ALMA Band 7 continuum map, with white contours superimposed 
on the \hst\ ACS F555M image from \citet{reines08}.
The black contours show the X-band Very Large Array (VLA) image (3.6\,cm) taken from \citet{johnson09}.
The 870\,\micron\ Band 7 emission is elongated in the east-west direction, and is slightly resolved
(there is a possible noise excess to the north-west).
The elliptical beam shape precludes a definitive analysis of the orientation
and extent of the emission.
In any case, like the cm-emission, the Band 7 emission is 
associated with the two brightest SSCs to the southeast.
The peak of the ALMA continuum lies closer to SSC\,1 (the southernmost
cluster) than SSC\,2,
although slightly offset from its optical peak. 

We could be missing some diffuse low-surface brightness emission
in \sbs\
because of the angular-size limitations of the extended ALMA configuration.
The field-of-view of our map is 17\arcsec\ (only a subset of the image is shown
in Fig. \ref{fig:sbsalma}), $\sim$5\,kpc, so clearly covers the entire
galaxy and even some extended \hi\ emission outside the optical confines.
Nevertheless, the nature of the ALMA interferometric measurement means
that emission at angular scales $\ga$3\arcsec\ could be missed;
such a spatial scale would be roughly the size of Fig. \ref{fig:sbsalma}
or larger.
Although we feel that this is unlikely in \sbs, additional flux at 870\,\micron\
would increase the dust mass we infer in Sect. \ref{sec:sed}.

\subsection{Free-free emission at 870\,\micron\ in \sbs}

870\,\micron\ emission can contain contributions from different
physical processes: cool dust, 
thermal free-free emission (or bremsstrahlung) from ionized gas,
and synchrotron emission from supernova remnants and 
their diffused non-thermal emission after longer timescales.
In \sbs, synchrotron emission contributes a significant fraction of the total
at long cm wavelengths and low spatial resolution \citep{hunt04}, 
but not at the highest radio frequencies observed \citep[43\,GHz,][]{johnson09}.
Extrapolating from the 1.3\,cm flux of $\sim$0.52\,mJy \citep{hunt04,johnson09}, we estimate a 
free-free flux at 870\,\micron\ of 0.40\,$\pm$\,0.07\,mJy.
We can also infer the free-free emission from the \bra\ flux \citep{hunt01},
although there may be some contribution from optically-thick winds in this
line \citep{hunt04,johnson09}.
From \bra, we obtain a similar ALMA Band 7 flux, 0.42\,$\pm$\,0.05\,mJy.
Taking the average of these, 0.41\,mJy,
we attribute $\sim$0.06\,mJy of the total ALMA Band 7 flux (0.47\,mJy)
to cool dust in \sbs.
By adding in quadrature the uncertainties in the Band 7 continuum level and
in the radio free-free flux,
we obtain the uncertainty in this dust-only 870\,\micron\ flux, 
$\sim$0.07\,mJy, implying a very marginal detection. 
Nevertheless, it gives us an idea of what to expect in the SED
fitting discussed in Sect. \ref{sec:sed}.
Taken at face value,
in \sbs\ at 870\,\micron, there is a factor of 7 
between the luminosity of the total emission including free-free,
and that of the dust; $\ga$87\% of the 870\,\micron\ emission is 
attributed to thermal free-free radiation.

\begin{figure*}[ht]
\centerline{
\includegraphics[angle=0,height=0.3\linewidth]{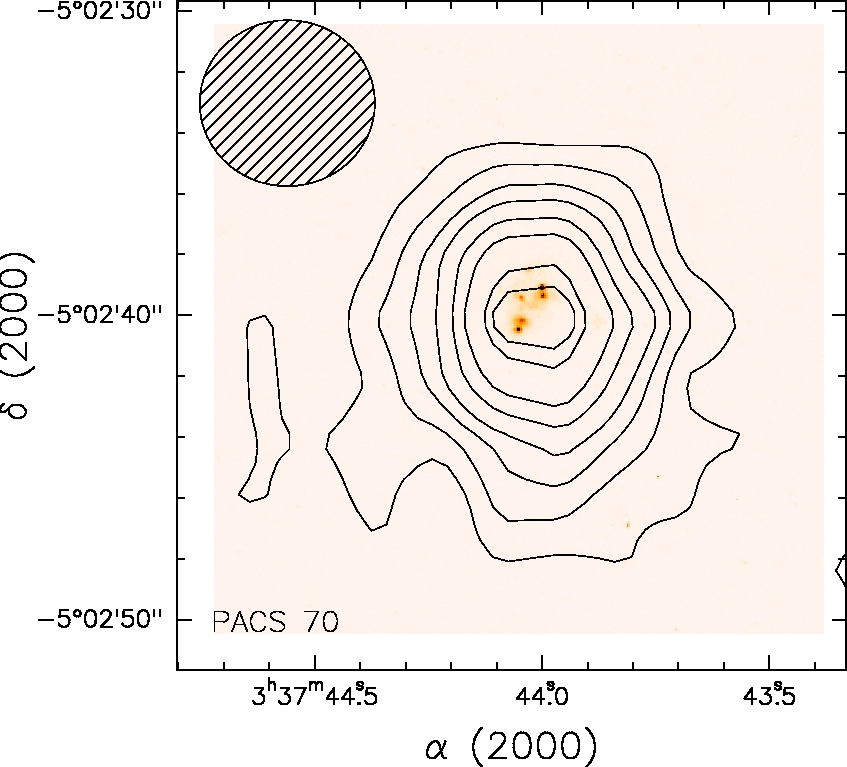}
\includegraphics[angle=0,height=0.3\linewidth]{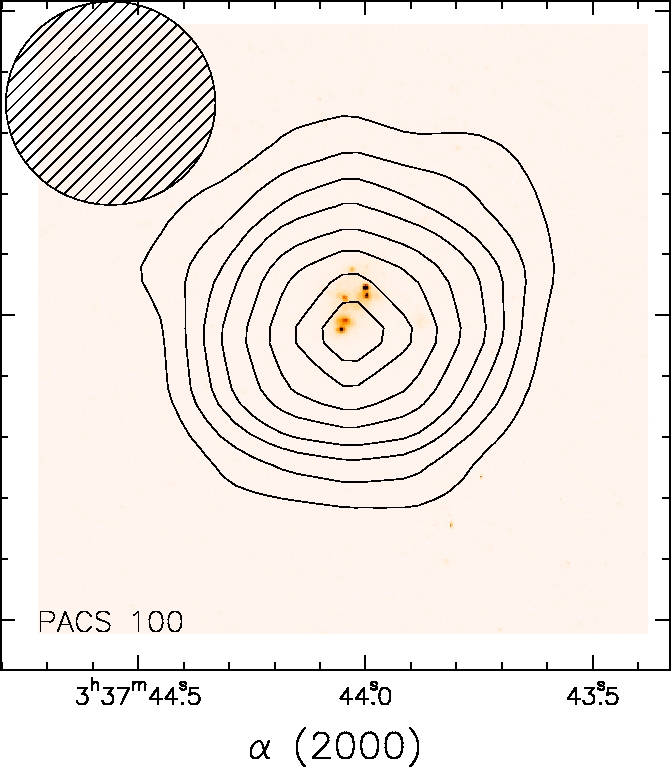}
\includegraphics[angle=0,height=0.3\linewidth]{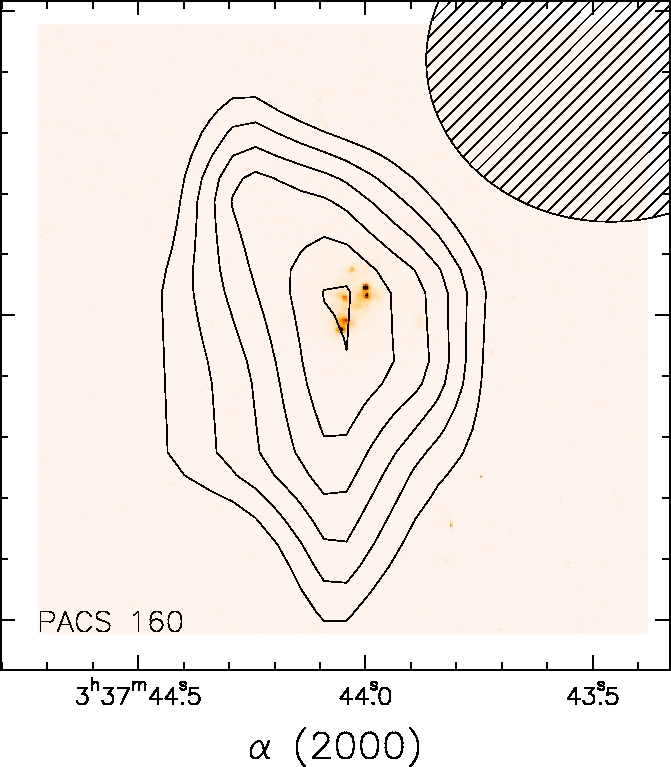}
}
\caption{PACS images of \sbs\ superimposed as contours
on the F555M HST/ACS image shown with false colors (see also Fig. \ref{fig:sbsalma}).
From left to right are PACS 70, 100, and 160\,\micron\ maps.
Contours start at 3$\sigma$ and run to 22$\sigma$ for PACS\,70,
to 20$\sigma$ for PACS\,100, and for
PACS\,160 to 8$\sigma$. 
These $\sigma$ values correspond to the correlated noise measured from the images, 
which are roughly 4$\times$ smaller than the true noise (see PACS documentation
and text).
PACS beam sizes are shown as shaded ellipses in the upper portion of each panel.
For display purposes, the PACS images have been rebinned to
smaller pixel sizes.
}
\label{fig:pacssbs}
\end{figure*}

\subsection{Herschel observations}
\label{sec:herschel}

We retrieved from the \hers\ Science Archive\footnote{
Herschel is an ESA space observatory with science instruments provided by 
European-led Principal Investigator consortia and with important participation from NASA.} 
PACS (Photodetector Array Camera 
\& Spectrometer\footnote{PACS has been developed by a consortium of institutes led by MPE
(Germany) and including UVIE (Austria); KU Leuven, CSL, IMEC (Belgium); CEA,
LAM (France); MPIA (Germany); INAF-IFSI/OAA/OAP/OAT, LENS, SISSA (Italy); IAC
(Spain). This development has been supported by the funding agencies BMVIT
(Austria), ESA-PRODEX (Belgium), CEA/CNES (France), DLR (Germany), ASI/INAF
(Italy), and CICYT/MCYT (Spain).}) 
images of \sbs, acquired in
Guaranteed Time with proposal KPGT\_smadde01\_1
(ObsID 1342202302, 134220303, 134220304, 134220305,
1342237748, 134223749).
PACS \citep{poglitsch10} observed \sbs\ at 70, 100, and 160\,\micron,
in large-scan mode during two visits. 
In the first visit, observations were acquired
with 10 legs in each scan map, of 3\arcmin\ length, separated by 4\arcsec,
repetition factor 1 which gives a total on-source exposure time
of 60\,sec. 
In the second visit, observations were acquired with scan legs of 
2\arcmin\ length, repetition factor 19, for a total integration time of 1140\,sec.
In both visits at each wavelength, in order to mitigate instrumental artefacts,
two different maps were acquired at 110$^\circ$ and 70$^\circ$, 
as recommended by the PACS instrument team. 
We reduced the observations with HIPE v.10.0 \citep{ott10}, starting from the pipeline reprocessed 
Level 0.5 data. 
The ``deep survey point-source" option was used, with masking performed on the images
themselves before combining the repetitions and orthogonal scans into a single map. 
In the end, for the final maps at 100 and 160\,\micron,
we used only the observations of the second visit, because of their superior
signal-to-noise;
70\,\micron\ maps were obtained only in the first visit.
Figure \ref{fig:pacssbs} shows the three PACS maps superimposed on the
\hst/ACS F555M image as in Fig. \ref{fig:sbsalma}.

We performed photometry in apertures between radius 10\arcsec\ and 14\arcsec,
and found that the flux remains stable to within a few percent in the largest
apertures; however, the noise grows with aperture size, so we used the values
at the 10\arcsec\ radius. 
Following the PACS calibration guidelines\footnote{These are found
at the URL: http://herschel.esac.esa.int/twiki/
bin/view/Public/PacsCalibrationWeb\#PACS\_calibration\_and\_performance.},
we adjusted the photometry for the appropriate aperture
correction, and corrected the uncertainty estimates for correlated noise. 
Color corrections are of order unity, and within the uncertainties, so we
neglected them.
Uncertainties on the photometry are estimated to be $\sim$8$-$9\% for PACS 70 and PACS 100,
including calibration and uncertainties on the sky subtraction;
the source is clearly detected at 70 and 100\,\micron. 
At 160\,\micron, the detection is roughly at 2.9$\sigma$.
Our photometry is slightly higher than, but
compatible within the errors to that reported by \citet{remy13}.

In order to search for extended emission,
we also reduced each of the PACS images with scanamorphos \citep{roussel13},
starting from the pipeline reprocessed Level 0.5 data.
The photometry of the resulting images gave in fact slightly lower fluxes
with larger uncertainty, perhaps because of the slightly larger reconstructed beams.
Hence, we adopt the photometry as described above, and are fairly confident
that the reduction technique did not miss any extended emission at any of the
PACS wavelengths.

\subsection{Other multi-wavelength data}
\label{sec:otherdata}

The broadband optical, near-, and mid-infrared portions of the SED were taken
from the literature \citep{papaderos98,vanzi00,dale01},
including data from \spit/IRAC \citep[InfraRed Array Camera,][]{fazio04},
\spit/IRS \citep[InfraRed Spectrograph,][]{houck04b}, and
\spit/MIPS \citep[Multiband Imaging Photometer,][]{rieke04},
taken from \citet{houck04a} and \citet{engelbracht08}.
For use in the SED fitting,
emission features (fine-structure lines) 
were subtracted from the \spit/IRS spectrum \citep{houck04a},
and the spectrum was then averaged over 0.5\,\micron\ bins.

Photometry from 1 to 10\,\micron\
was corrected for nebular continuum
emission based on the star-formation rate (SFR),
as described in \citet{hunt12}.
The SFR was converted to free-free emission by first using the conversion
factor of \citet{kennicutt98} to estimate \ha\ luminosity, then  
with the coefficients in \citet{osterbrock06} to \bra\ and 
free-free and free-bound continuum levels.
We compared this method based on the SFR with
the directly observed \bra\ flux \citep{hunt01}
and the high-frequency radio emission \citep{hunt04,johnson09} and
obtain consistent results.
This is an important correction since
the free-free contribution at 2\,\micron\ is 55\%, and $\sim$20\%
at 3.6 and 4.5\,\micron.
We also included in the SED the cm-wavelength radio data from \citet{hunt04} and
from \citet{johnson09}.
Table \ref{tab:sedsbs} reports the photometry described above, corrected for
nebular continuum where noted.

%
\begin{center}
\begin{table}
      \caption[]{Global photometry for \sbs} 
\label{tab:sedsbs}
\addtolength{\tabcolsep}{3pt}
{
\tiny
\begin{tabular}{lllc}
\hline
& & \multicolumn{1}{c}{1$\sigma$}\\
\multicolumn{1}{c}{Wavelength}  & \multicolumn{1}{c}{Flux$^{\mathrm{a}}$} & 
\multicolumn{1}{c}{Uncertainty} & Reference$^{\mathrm{b}}$ \\
\multicolumn{1}{c}{(\micron)}  & \multicolumn{1}{c}{(mJy)} & 
\multicolumn{1}{c}{(mJy)} \\
\hline
0.360 &    0.73  &  0.021  & (1) \\ 
0.440 &    0.85  &  0.013  & (1) \\ 
0.550 &    0.92  &  0.015  & (1) \\ 
0.641 &    0.81  &  0.007  & (1) \\ 
0.791 &    0.49  &  0.004  & (1) \\ 
1.25 &    0.40$^\mathrm{c}$  &  0.023  & (2) \\ 
1.65 &    0.25$^\mathrm{c}$  &  0.017  & (2) \\ 
2.20 &    0.31$^\mathrm{c}$  &  0.0204 & (2) \\ 
3.55 &    0.62$^\mathrm{c}$  &  0.0234 & (3) \\  
4.49 &    1.50$^\mathrm{c}$  &  0.0493 & (3) \\  
5.73 &    4.07$^\mathrm{c}$  &  0.1320   & (3) \\
7.87 &    12.73$^\mathrm{c}$ &   0.3890  & (3) \\ 
12.3   &    35    &     6     & (4) \\   
23.7   &      79.0 & 3.12    & (3) \\
71.1   &   64.4   &  5.7    & This paper (PACS$^{\mathrm d}$)\\ 
71.4   &      52.4 & 4.75    & (3) \\
101.2  &   31.3   &  2.7    & This paper (PACS$^{\mathrm d}$)\\ 
162.7  &   10.4   &  3.5    & This paper (PACS$^{\mathrm d}$)\\ 
866.5  &   0.47$^\mathrm{e}$   & 0.035   & This paper\\ 
866.5  &   0.07$^\mathrm{f}$   & 0.070   & This paper\\ 
13324  &     0.48 &   0.0789 & (5) \\  
20120  &     0.40 &   0.0704 & (5) \\  
35436  &     0.65 &   0.0386 & (5) \\  
61685  &     0.77 &   0.0855 & (5) \\  
205337 &    0.46  &  0.0610  & (5) \\  
13324  &    0.56  &  0.0789  & (6) \\  
20120  &    0.58  &  0.0704  & (6) \\  
35436  &    0.46  &  0.0386  & (6) \\  
61685  &    0.34  &  0.0855  & (6) \\  
\hline
\end{tabular}
}
\vspace{0.5\baselineskip}
\begin{description}
\item
[$^{\mathrm{a}}$] 
Fluxes given here are corrected for Galactic extinction, \av\,=\,0.155\,mag.
\item
[$^{\mathrm{b}}$] 
(1) \citet{papaderos98};
(2) \citet{vanzi00};
(3) \citet{engelbracht08};
(4) \citet{dale01};
(5) \citet{hunt04};
(6) \citet{johnson09}.
\item
[$^{\mathrm{c}}$]
Corrected for nebular continuum as described in the text.
\item
[$^{\mathrm{d}}$] 
Photometry in an aperture of 10\,\arcsec\ radius.
\item
[$^{\mathrm{e}}$] 
Total ALMA Band 7 continuum emission.
\item
[$^{\mathrm{f}}$] 
Band 7 dust emission only, corrected for free-free contamination
as described in the text.
\end{description}
\end{table}
\end{center}

\section{The spectral energy distribution of SBS0335$-$052}
\label{sec:sed}

We have used the multi-wavelength dataset in Table \ref{tab:sedsbs}
to estimate the dust mass in \sbs\
by fitting a spherical \dusty\ model \citep{ivezic97}.
Before fitting the SED, photometry was corrected for Galactic extinction
assuming \av\,=\,0.155\,mag (taken from the NASA Extragalactic Database,
NED, {\tt http://ned.ipac.caltech.edu}), and using the interstellar
extinction curve by \citet{cardelli89}.
Because of the relatively large size of the IR beams (\spit, \hers), 
we have considered the global photometry rather than that for only
SSCs 1$+$2; 
although the optical \hst\ data would have allowed
such a separation \citep[e.g.,][]{reines08}, in the IR such a separation would be virtually 
impossible. 

The \dusty\ formalism solves analytically the radiative transfer problem 
of a spherical dust shell heated by a point-like heating source 
whose immediate vicinity is devoid of dust.
Thus, our model consists of two SED components: the stellar heating spectrum
and a \dusty\ spherical shell surrounding the internal radiation
source. The dust re-processes a certain fraction of the
incident radiation into the IR regime, and produces the blackbody-like
peak in the total SED. The total SED is thus modeled as
\begin{equation}
  \label{eq:model}
  {\rm SED}_\lambda = a\ {\rm HFN}_\lambda \langle \exp (
  -\tau_\lambda^{{\rm stars}})\rangle \, +\, b\ {\rm IR}_\lambda \ ,
\label{eqn:dusty}
\end{equation}
where the total SED, the heating function HFN, and the IR emission are
wavelength-dependent. 
The stars comprising the HFN will have different obscuring $\tau(V)$
toward them, so we use the average of a mix of stars and dust: 

\begin{equation}
\langle I_0\,e^{-\tau_\lambda} \rangle\,=\,\langle I_0 \rangle \langle e^{-\tau_\lambda} \rangle \ ,
\end{equation}
where
\begin{equation}
\langle e^{-\tau_\lambda} \rangle\,=\,\frac{1}{\tau_\lambda^{\rm max}} \int^{\tau_\lambda^{\rm max}}_0
e^{-\tau_\lambda}\,d\tau \,=\,\frac{1-e^{-\tau_\lambda^{\rm max}}} {\tau_\lambda^{\rm max}} \ .
\end{equation}
The stellar component is thus attenuated by a
foreground screen of this average ($V$-band) optical depth 
$\langle \tau_\lambda^{\rm stars}\rangle$,
not necessarily the same as that of the dust itself. 

$a$ and $b$ 
are independently determined scales for
the two components of the total SED, such that together they maximize the
likelihood of the model fitting the observations. 
The reason behind two different normalizations is that
\dusty\ assumes a uniform dust cloud with a filling factor of unity.
Hence, in our fits,
the same \dusty\ template is used to fit both stars and dust, but the relative
normalization is allowed to vary in order to accommodate non-unity values
of the fraction of dust that intercepts stellar light.
The optical depth of the dust that reprocesses stellar radiation can be
different from the optical depth of the dust that emits directly because
dust may be clumpy, with the clumps having a higher opacity
than the (possibly more diffuse) dust that absorbs stellar light.
Once the best-fit parameters are defined, the 
shapes of all SED components are known, and the normalization can be found
analytically \citep{nikutta12}.

The optical regime of the \dusty\ models requires three parameters:
dust optical depth (defined in the $V$ band) for the stars, stellar age and metallicity.
The input parameters for the IR portion of the \dusty\ models
are:
the temperature at the inner part of the dust shell, \tin; 
the ratio of the outer shell radius to that of the inner shell boundary,
$Y\,\equiv\,R_{\rm out}/R_{\rm in}$;
the power law index $p$ of the dust distribution $\propto R^{-p}$;
and the $V$-band dust optical depth \tauvd.
The \dusty\ template library includes
HFNs modeled as the single stellar population models of \citet{bc03}
at three different metallicities (0.02\,\zsun, 0.1\,\zsun, and \zsun)
and 12 different ages from 1\,Myr to 10\,Gyr,
assuming a \citet{chabrier03} Initial Mass Function.
Also included in the library are three different dust grain populations:
Milky-Way dust \citep{draine07},
dust in the Small Magellanic Cloud (SMC) \citep{weingartner01},
and chemically unevolved dust in primordial supernovae (SNe)
\citep{bianchi07}.

We find the best fits and the range of model parameter values
compatible with the data via exploration of the seven-dimensional
(7D) parameter space with an efficient Markov Chain Monte Carlo 
code\footnote{We based it on the PyMC package:
\url{http://pymc-devs.github.io/pymc/}}. 
The three different dust-grain populations are fitted separately,
and results are compared a posteriori.
The HFNs and IR SEDs
are tabulated on a 7D-rectilinear parameter grid, 
then interpolated for the specific trial parameters using
the implementation from \citet{nikutta12}. 
We calculate the likelihood of each model
assuming Gaussian flux errors on the measurements. 
The absolute
normalization of the total posterior is irrelevant for the parameter
estimation problem.

For most of the parameters to be fit, we assumed uniform prior probability
density functions.
However, 
we constrained the metallicity to be close to that measured by
\citet{izotov99}, 
$0.0347\,\pm\,0.014$\,Z$_\odot$. 
We  imposed a truncated Gaussian as a prior, centered at 0.0347\,\msun, with
a width equal to the $1\sigma$ uncertainty (truncated at the $1\sigma$ limits).
The final fit gives the best-fit parameters 
(the normalized marginalized posteriors) and their confidence intervals.

\subsection{SED-fitting results}

\begin{figure}[ht]
\centerline{
\includegraphics[angle=0,width=\linewidth]{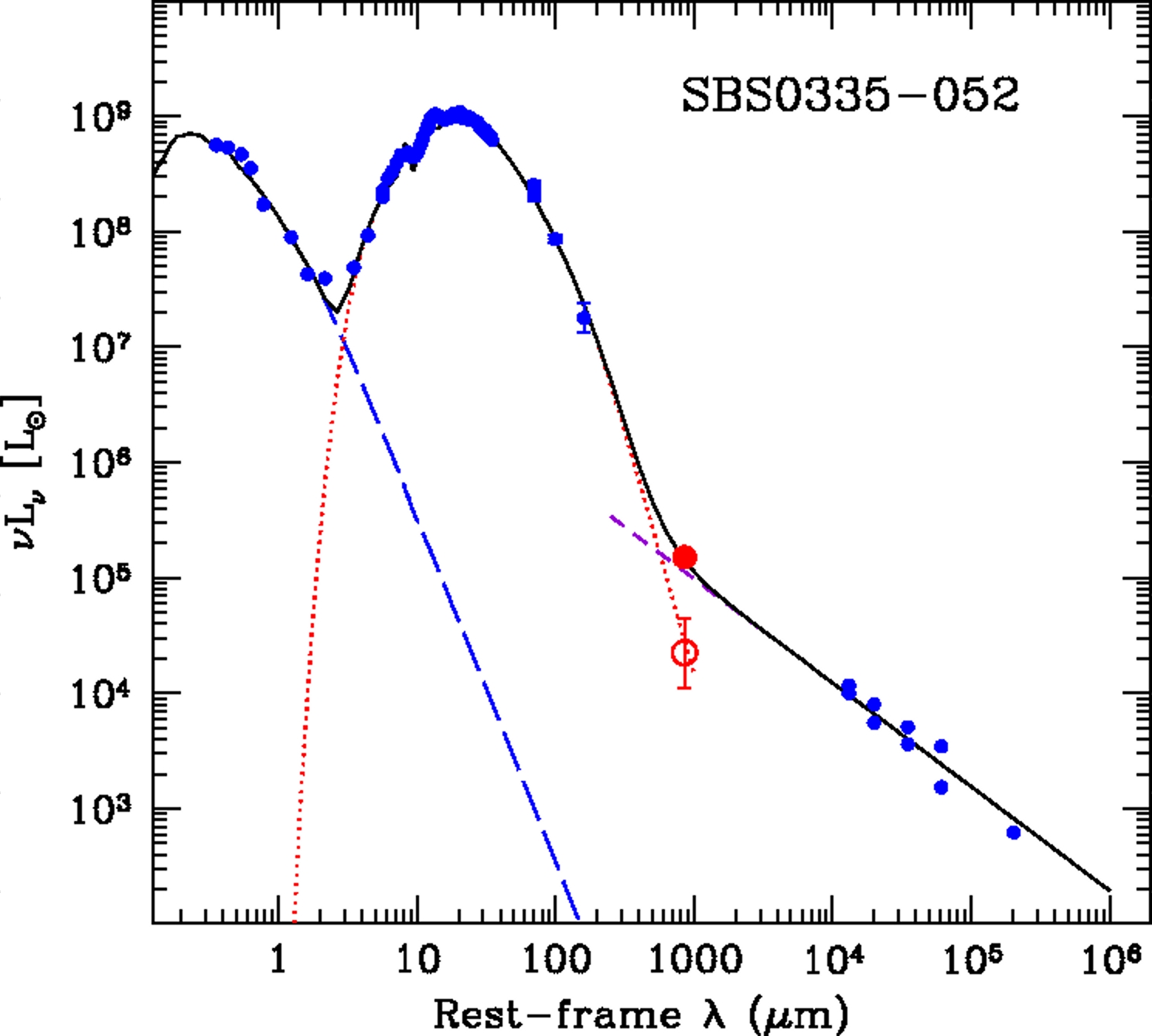}
}
\caption{SED of \sbs, together with best-fit \dusty\ model overlaid.
The two large (red) circles show the total ALMA Band 7 emission;
the dust-only emission, estimated as described in the text, is indicated
by the open (red) circle. 
The black curve represents the \dusty\ fit, and the dotted (red) 
curve, only the dust; the long-dashed (blue) curve shows the stars.
The radio slope is pure free-free, $\propto\nu^{-0.1}$, and the
pure radio component is shown by a short-dashed (purple) line
(longward of 250\,\micron).
\dusty\ models only extend to 1\,mm, so the excess above the dust-only
emission is the extrapolation of the radio. 
When not evident, the error bars are masked by the size of the symbols.
}
\label{fig:sedsbs}
\end{figure}

Figure \ref{fig:sedsbs} shows the best-fit \dusty\ model for the SED of \sbs.
The radio emission has been approximated using a spectral index of
$\alpha=-0.1$, appropriate for pure thermal emission.
The radio emission is generally well fit by such a spectrum, although
at low frequencies the spectrum is self-absorbed by the dense gas
in the radio nebulae and the synchrotron component
begins to be significant \citep{hunt04,johnson09}.

The SED of \sbs\ is very warm;
as already noted by \citet{houck04a},
the SED peaks between 20 and 30\,\micron, even more extreme than the 60\,\micron -peaker
galaxy population discovered with the Infrared Astronomical Satellite (IRAS) \citep{vader93}.
The temperature at the inner radius of the shell is 
497$^{+5}_{-14}$\,K; this relatively high
temperature is necessary to fit the rapidly-rising mid-infrared portion of the spectrum.
This short-wavelength peak combined with the intense star-forming event
makes it difficult to detect 
the cool dust in the SEDs of metal-poor dwarf starbursts.

The best-fit stellar population age is 13.1$^{+2.8}_{-0.5}$\,Myr, 
at the lowest metallicity sampled (2\% \zsun).
This age is consistent with the oldest of the SSC ages estimated by \citet{reines08}.
In fact,
there is an age gradient in the clusters that could be as large
as 25\,Myr from the oldest (toward the northwest) to the youngest
\citep[in the southeast,][]{thuan97b}, or even larger \citep{papaderos98}.
Thus, an age of 13\,Myr seems like a reasonable compromise for our global photometry
which is a luminosity-weighted average.
The stars themselves suffer from a total extinction $\tau^{\rm stars}_V$\,=\,0.5$^{+0.2}_{-2.5}$,
significantly smaller than the optical depth of the dust clouds (see below).
The stellar mass can be inferred from the \dusty\ fit, and we find
\mstar\,=\,$2.35\times10^7$\,\msun, 
from the $V$-band luminosity of $5.6\times10^8$\,\lsun\ (corrected
for extinction) and assuming a $V$-band mass-to-light ratio
($M/L_V$) of 0.04 \citep{bc03}, appropriate for a 13\,Myr stellar population.
This is 4 times greater than the mass given by 
\citet{reines08},
obtained by fitting broadband optical-near-infrared SEDs 
of the 6 individual SSCs with single stellar population models
\citep[see also][]{fumagalli10}.
However, the age found by \dusty\ corresponds to the oldest cluster,
which would have a $M/L_V$ ratio 4 times larger than a 4\,Myr population.
When we impose a prior on the stellar age (in the range of 3 to 6\,Myr),
the \dusty\ fit returns a best-fit age of 4\,Myr and a
$V$-band luminosity of $6.8\times10^8$\,\lsun\ (also corrected for extinction).
With $M/L_V$\,=\,0.01, appropriate for 4\,Myr, we find a 
stellar mass of $6.8\times10^6$\,\msun,
only 20\% more than the Reines et al. value.
We thus conclude that the \dusty\ stellar mass is roughly consistent
with earlier estimates.

\begin{figure*}[ht]
\centerline{
\includegraphics[angle=0,height=0.35\linewidth]{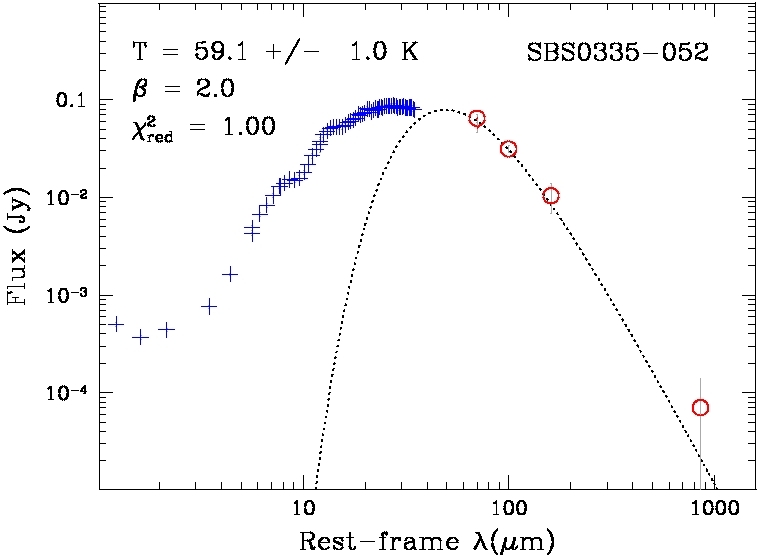}
\includegraphics[angle=0,height=0.35\linewidth]{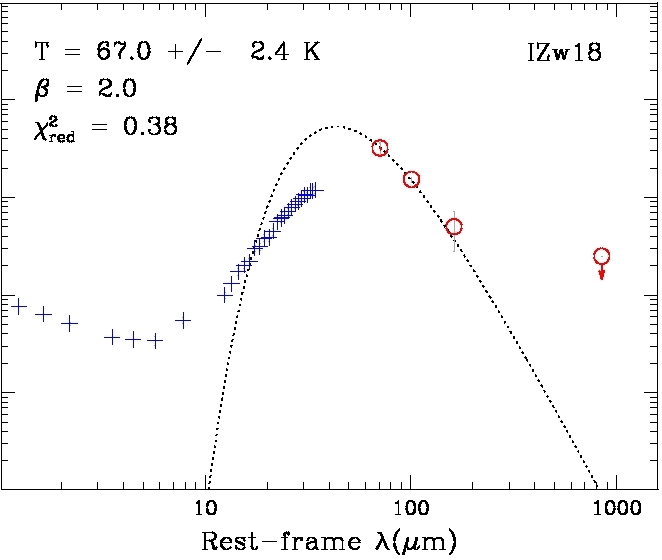}
}
\caption{Modified single-temperature blackbody fits with fixed $\beta$
of \sbs\ (left panel) and \izw\ (right panel).
The blue $+$ show unfitted points; the upper limit at 850\,\micron\ for \izw\ was also not fit.
In the right panel, the MIPS and PACS \,70\,\micron\ points are so close in wavelength
and measured flux that they are virtually indistinguishable.
Details for \izw\ are given in Sect. \ref{sec:izw}.
}
\label{fig:MBB}
\end{figure*}

Because there is a clear silicate absorption feature at 9.7\,\micron\
\citep{houck04a}, the dust must contain silicates;
SMC dust is the grain population that provides the best fit. 
To reproduce this feature in absorption, it is necessary to have a relatively large
quantity of cool dust in front of the massive stars heating the dust.
This is naturally achieved in \dusty\ through a large dust optical depth,
$\tau_{\rm dust}$, with a relatively thick shell and
a large power-law index. 
In the case of \sbs, the best-fit \tauvd\,=\,11.8$^{+0.6}_{-2.5}$; 
the shell relative thickness $Y\,=\,R_{\rm out}/R_{\rm in}$ is 
605$^{+58}_{-21}$,
and the power-law index of the radial dust distribution ($\propto R^{-p}$), $p$\,=\,1.63$^{+0.07}_{-0.03}$.
With a thinner shell and a less steep (more uniform) radial dust distribution, 
the silicate feature would emerge in emission rather than in absorption.

From the best-fit \dusty\ model we obtain the dust mass, since the geometry
of the dust and the optical properties of the grains are known.
We find that there are $(3.8\,\pm\,0.6)\times10^4$\,\msun\ of dust in \sbs.
The normalization of the \dusty\ models also gives the diameter of the 
dust-emitting region, 2\farcs4 or $\sim$660\,pc, 
comparable to the region ($\sim$2\farcs1)
emitting faint \pa\ \citep{reines08}.

We performed two ``sanity checks'' for the derived dust mass: first, we performed a
single-temperature modified blackbody fit (MBB) with variable emissivity, $F_\nu\propto\nu^{\beta}\,B(T,\nu)$,
where $B(T,\nu)$ is the Planck function.
The MBB fitting was performed by fixing $\beta$ to 2
because otherwise dust mass estimates
are unreliable due to the dust opacities which are calibrated with roughly
this value \citep{bianchi13}.
Allowed temperatures ranged from 5\,K to 100\,K because color corrections for PACS and
MIPS are well defined only over this temperature range \citep[e.g.,][]{aniano12}.
This fit\footnote{This fit has a $\chi^2_\nu$
value (1.00) not significantly worse than that performed by letting also $\beta$ vary (0.86).},
shown in Fig. \ref{fig:MBB}, results in a temperature $T$\,=\,59$\pm$\,1\,K.
Using this fit at 100\,\micron\, and assuming the SMC emissivity, we find
a dust mass of $3.4\times10^3$\,\msun, a factor of 10 lower than the \dusty\ value.
However, Fig. \ref{fig:MBB} shows that the single-$T$ MBB fit misses a large
fraction of the warmer dust, and does not pass through the ALMA point so it
is very likely that a significant fraction of cooler dust is also missing.
We thus conclude that the \dusty\ value is not unreasonable.

As a second check,
we compared the \dusty\ mass with that inferred from the
dust optical depth at 870\,\micron. 
Assuming the mean dust temperature of $\sim$59\,K given by the MBB fit, 
and a mass coefficient at 870\,\micron\ for SMC dust of 
$\kappa_\nu$\,=\,4.5\,m$^{2}$\,kg$^{-1}$
\citep{weingartner01}, we obtain a dust mass surface density of
\sigmadust\,=\,0.45\,\msunpc.
The integrated area of the region with ALMA Band 7 emission $\geq3\sigma$ is
$\sim$1\,arcsec$^2$.
We would thus estimate a total dust mass of $3.1\times10^4$\,\msun,
well within the uncertainties of the total dust mass given by 
the \dusty\ fit.
The dust mass surface density of $\sim$0.75\,\msunpc\
inferred from the silicate optical depth $\tau_{9.7}$ 
\citep[$\sim$0.46,][]{houck04a} is higher than that from the 870\,\micron\ dust emission.
This is perhaps
not surprising because the dust traced at 9.7\,\micron\
in absorption is expected to be more concentrated \citep[e.g.,][]{dale01}.

\begin{figure*}[ht]
\centerline{
\includegraphics[angle=0,height=0.35\linewidth]{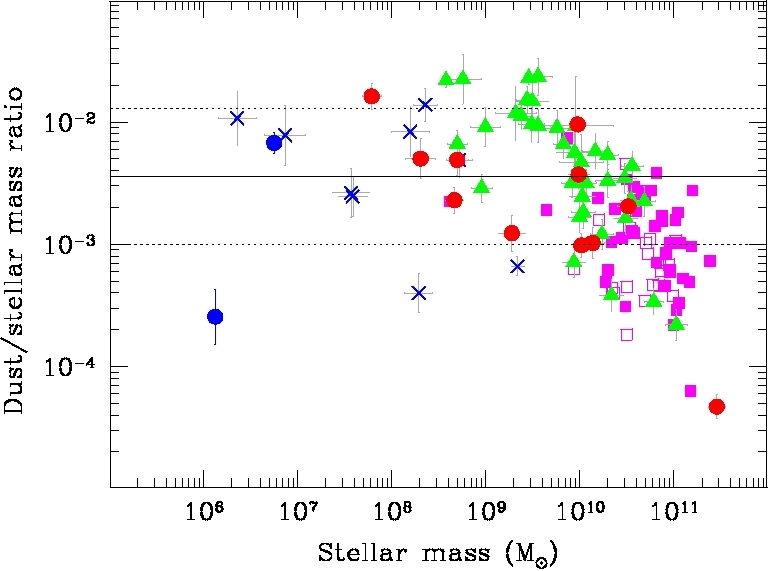}
\includegraphics[angle=0,height=0.35\linewidth]{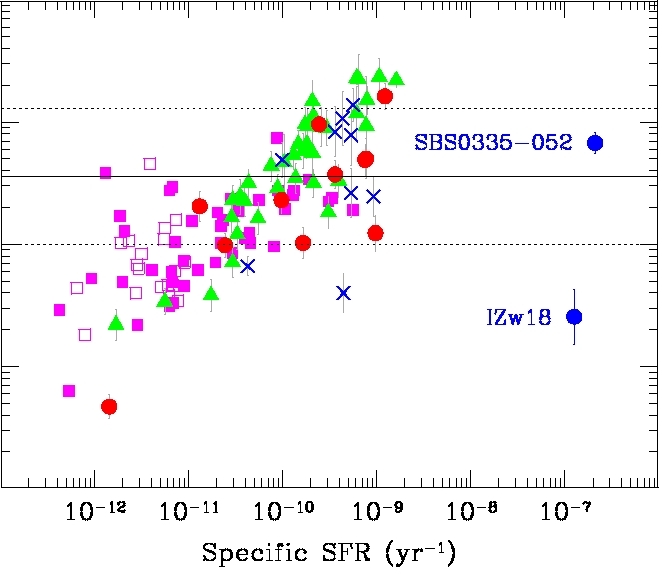}
}
\caption{Ratios of dust mass to stellar mass plotted against stellar mass (\mstar\ left panel)
and specific SFR (yr$^{-1}$, right).
\sbs\ and \izw\ are plotted as filled (blue) circles.
Data for KINGFISH galaxies, taken from \citet{kennicutt11}, are also shown:
$\times$ (for late Hubble types T$\geq$8),
filled (green) triangles (0$<$T$<$8), and filled (red) circles (for early Hubble types T$\leq$0).
The horizontal lines give the mean dust/stellar mass ratio for 
the KINGFISH sample, and the 1$\sigma$ standard deviation.
Given as open (magenta) squares are the passive spirals studied by \citet{rowlands12}
and as filled (magenta) squares their early-type galaxies.
}
\label{fig:ratio}
\end{figure*}

Our estimate of the dust mass in \sbs\ is more than 40 times higher than the value
of 800\,\msun\ given by \citet{remy13}.
There are several reasons for this.
The first, most obvious one, is that they use an 
emissivity at 100\,\micron\ $\kappa_\nu$\,=\,4.5\,m$^{2}$\,kg$^{-1}$,
$\sim$30\,\% higher than the best-fit SMC dust with $\kappa_\nu\,=\,$3.47\,m$^{2}$\,kg$^{-1}$ \citep{weingartner01}.
With this emissivity, \citet{remy13} would have obtained 1037\,\msun, rather than 800\,\msun.
Second, they included the 24\,\micron\ point in their single-temperature
MBB fit which skews fitted temperatures to higher values, thus lowering
the inferred dust mass.
Indeed, if we repeat our MBB fit to the fluxes in Table \ref{tab:sedsbs} letting
$\beta$ vary and including the 24\,\micron\ point, we obtain $T\,=\,87.5$\,K, 
$\beta\,=\,$1.2, 
similar to $T\,=\,89^{+10}_{-8}$\,K and $\beta\,=\,1.64^{+0.39}_{-0.33}$
given by \citet{remy13}.
Our MBB fit including the 24\,\micron\ point would correspond to a dust mass
of $1.4\times10^3$\,\msun, only 35\% higher than the value of \citet{remy13} with
the SMC emissivity, but
less than half the value we obtain without including the 24\,\micron\ point.
The inclusion of the 24\,\micron\ point leads to lower dust masses because of
the invalidity of the assumption of a single temperature
\citep[e.g.,][]{dale12}. 
In the radial temperature profile given by the \dusty\ fit of \sbs, 
dust temperatures fall to $\sim$20\,K
in the regions farthest from the radiation source 
even though the innermost dust is very warm, $\sim$500\,K.

A third reason for the difference in dust masses
is that the dust in \sbs\ is not optically thin at short
wavelengths, as implicitly assumed in the use of MBB fits.
The clear silicate absorption feature and the behavior of the SED as modeled
by \dusty\ are signatures of the optical thickness of the dust radiation
and the importance of radiative transfer.
A fourth reason is that by letting $\beta$ vary in the MBB fits as done by
\citet{remy13}, long-wavelength constraints become important.
When we fix $\beta$ to 1.6 as in \citet{remy13}, 
we obtain roughly the same value of dust mass as for $\beta\,=\,1.2$
(our best-fit $\beta$ value, see above).
When we let $\beta$ vary in the MBB fitting,
our data result in a flatter $\beta$ slope because of the 
added constraint of 870\,\micron.
On the other hand, when $\beta$ is fixed, 
the need to accommodate longer-wavelength constraints with 
steeper (larger values of) $\beta$ will lead to lower temperatures 
and consequently larger dust masses 
\citep[e.g.,][]{bianchi13}.

The dust mass from the best-fit \dusty\ model, $3.8\times10^4$\,\msun,
gives a dust-to-stellar mass ratio of $\sim$0.007
\citep[assuming the stellar mass from][see above]{reines08}.
Ratios of dust-to-stellar mass for \sbs\ are compared in Fig. \ref{fig:ratio}
to the KINGFISH \citep{kennicutt11} sample of galaxies
and to passive spirals and early-type galaxies from H-ATLAS/GAMA \citep{rowlands12}.
Also shown is \izw\ (see Sect. \ref{sec:izw}).
H-ATLAS/GAMA values are derived from optical-infrared SED fitting.
For KINGFISH,
stellar parameters are taken from \citet{kennicutt11}, or when not available, from
\citet{skibba11} rescaled to the correct distances; dust masses are derived by \citet{bianchi13}.
Since in these galaxies dust is usually spatially coincident with the stars, we have used
global values for both sets of masses. 
Figure \ref{fig:ratio} shows that there is considerable scatter
at low stellar masses, \mstar, 
but for more massive galaxies with \mstar$\ga 10^{10}$\,\msun,
dust-to-stellar mass ratios tend to be lower than at the low-mass end.
The ratio for \sbs\ is within the range of the KINGFISH galaxies
\citep[see also][]{cortese12}.

The right panel of Fig. \ref{fig:ratio} shows the dust-to-stellar mass ratios plotted against
specific SFR, the ratio of SFR to \mstar\ (sSFR).
As found by \citet{dacunha10}, these two quantities are well correlated.
Such a correlation would be expected if dust is broadly tracing the ISM content 
\citep[i.e., mostly gas with approximately the same dust-to-gas ratio, see for example][]{eales12}. 
Galaxies with higher sSFR would be in 
earlier evolutionary stages with a higher ISM fraction.
\sbs\ has a sSFR $\sim$100 times higher than the highest sSFR in the KINGFISH
sample which implies that it is in a very early phase of its evolution. 
The comparison with the KINGFISH and H-ATLAS/GAMA samples shows that the galaxies with
the highest sSFRs, including \sbs, generally 
have high dust-to-stellar mass ratios.
While it would be tempting to draw a regression line between dust-to-stellar
mass ratios and sSFR, models suggest that for sSFR\,$\ga 10^{-9}$\,yr$^{-1}$, 
the dust content falls \citep[e.g.,][]{calura08,dacunha10}. 
The early evolutionary stages implied by high sSFRs 
would presumably be in the dust-formation phase of the ISM, 
implying that the overall dust content is lower relative to later more evolved stages
of evolution. 

\begin{figure}[ht]
\centerline{
\includegraphics[angle=0,width=\linewidth]{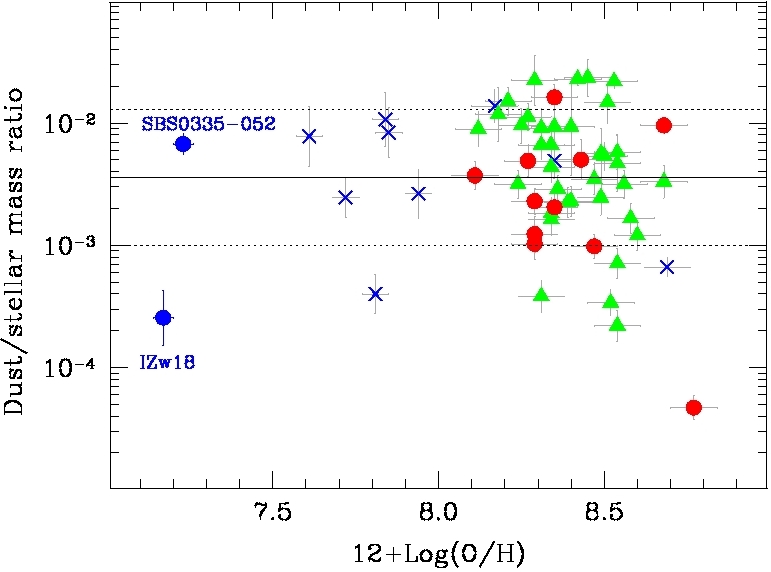}
}
\caption{Ratios of dust mass to stellar mass plotted against nebular \logoh.
\sbs\ and \izw\ are plotted as filled (blue) circles. 
As in Fig. \ref{fig:ratio},
data for KINGFISH galaxies (see text) are also shown:
$\times$ (for Hubble type T$\geq$8),
filled (green) triangles (0$<$T$<$8), and filled (red) circles (T$\leq$0).
The horizontal lines give the mean dust/stellar mass ratio for 
the KINGFISH sample, and the 1$\sigma$ standard deviation.
}
\label{fig:ratiooh}
\end{figure}

Figure \ref{fig:ratiooh} plots dust-to-stellar mass ratios vs. nebular oxygen abundance, 
\logoh\footnote{These are the \citet{pilyugin05} calibrations from \citet{kennicutt11}, in order to be
more comparable with the direct method used for \sbs\ and \izw\
based on electron temperature.}.
The mass ratios are less well correlated with metallicity than 
with stellar mass or sSFR. 
\sbs\ and \izw\ (see next section) have a similarly low metallicity,
being among the most metal-poor star-forming galaxies in the Local Universe, but their
dust-to-stellar ratios differ dramatically;
however both galaxies have ratios that 
are similar to those of galaxies more than 30 times more metal rich.
Metallicity may be less important than other parameters for driving
star-formation processes \citep[e.g.,][]{hunt12}.

\section{Comparison with I\,Zw\,18}
\label{sec:izw}

As outlined in the Introduction, we wish to compare \sbs\ 
with another very low-metallicity star-forming dwarf galaxy,
\izw.
Here we describe the data we have assembled for the SED of \izw, and
the results from fitting it
with the same \dusty\ algorithm used for \sbs.

\subsection{Herschel data for I\,Zw\,18}
\label{sec:herschelizw}

\begin{figure*}[ht]
\centerline{
\includegraphics[angle=0,height=0.3\linewidth]{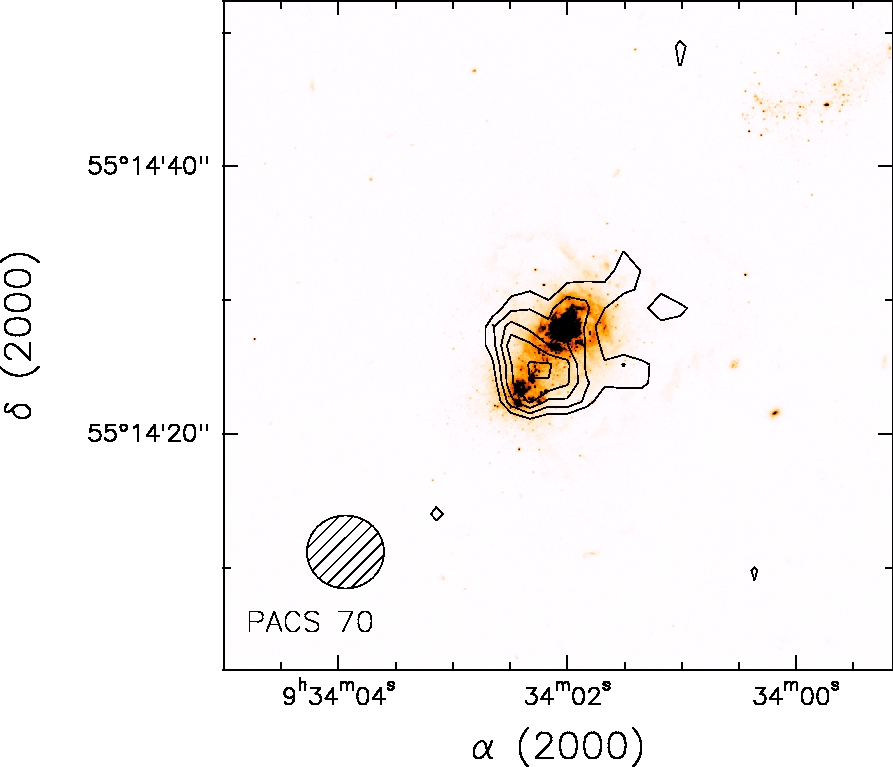}
\includegraphics[angle=0,height=0.3\linewidth]{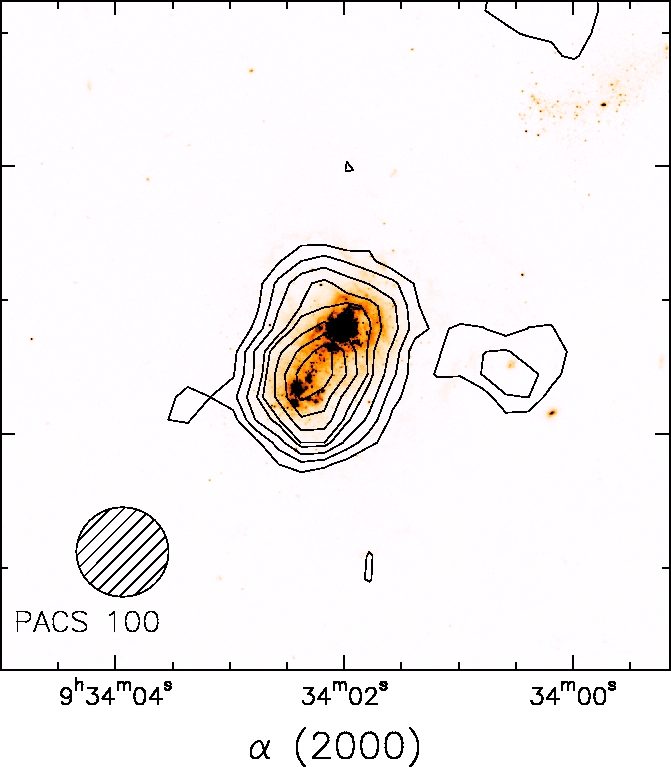}
\includegraphics[angle=0,height=0.3\linewidth]{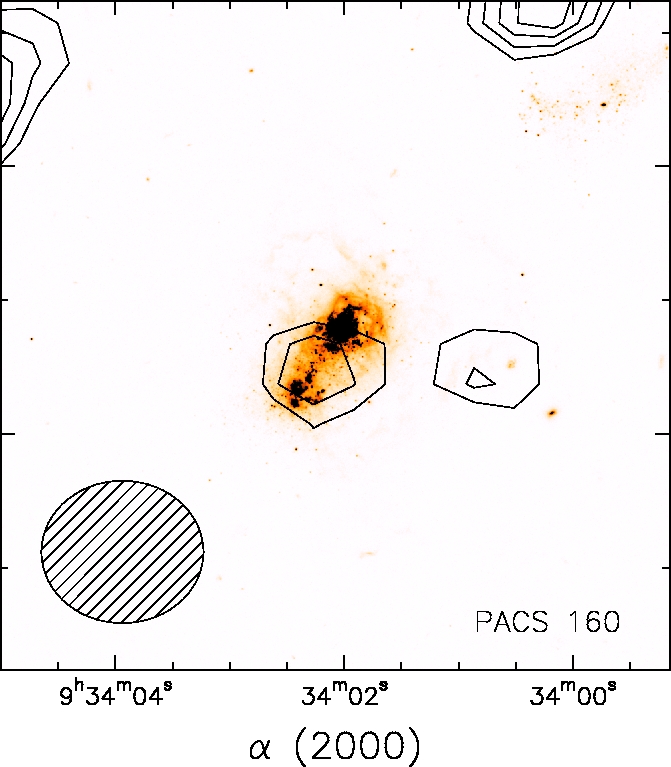}
}
\caption{PACS images of \izw\ superimposed as contours
on the F606W HST/ACS image shown with false colors.
From left to right are PACS 70, 100, and 160\,\micron.
Contours start at 3$\sigma$, and run to 7$\sigma$ for PACS\,70,
to 16$\sigma$ for PACS\,100, and for PACS\,160 to 2$\sigma$. 
As in Fig. \ref{fig:pacssbs},
these $\sigma$ values correspond to the correlated noise measured from the images, 
which are smaller than the true noise (see text). 
}
\label{fig:pacsizw}
\end{figure*}

We acquired from the \hers\ Science Archive
images of \izw, acquired in
Guaranteed Time with proposal KPGT\_smadde01\_1
(ObsID 1342209354, 1342209355, 1342209356, 1342209357)
and in Open Time with proposal
OT2\_dfisher\_1 (ObsID 1342245936, 1342245937).
For 100 and 160\,\micron, we used only the latter
observations as including also the former did not improve the noise characteristics
of the final images.
The data were reduced to Level 1 with HIPE 
v.10.0 \citep{ott10}, starting from the pipeline reprocessed 
Level 0.5 data, but because of the need for conserving extended
emission, we used scanamorphos \citep{roussel13} to create the final images
shown in Figure \ref{fig:pacsizw}. 

\subsection{I\,Zw\,18 photometry}

We also downloaded the \spit\ MIPS 24, 70,
and 160\,\micron\ datasets, and combined
the individual images into mosaics with MOPEX \citep{mopex}.
The emission in the 24\,\micron\ image extended to radii of 12-13\arcsec,
which dictated the apertures we used for photometry.
We used the 24\,\micron\ photometry to check if the IRS spectrum was missing
flux because of the $\sim$11\arcsec\ widths of the long-wavelength
slits.
After applying the MIPS aperture correction of 1.15, the measured
flux of 6.3\,mJy agrees very well with the 24\,\micron\ flux of
6.1-6.5\,mJy measured by IRS.

The MIPS and PACS photometry was performed over apertures of radius 13\arcsec, 
1.6 times the optical radius ($\sim$8\farcs1).
As for \sbs, we corrected the PACS photometry and uncertainties for
the aperture correction and correlated noise because of non-independent
pixels.
Color corrections are of order unity, so as for \sbs, we did not apply them. 
Uncertainties on the photometry are estimated to be 6-7\% for PACS 70
and $\sim$4\% for PACS 100,
including calibration and uncertainties of the sky subtraction.
For MIPS photometry, we applied aperture and color corrections as described
in the MIPS Data Reduction Guide.
Final uncertainties for MIPS including sky and calibration uncertainties 
are estimated to be $\sim$11\% at 24\,\micron, and $\sim$14\% at 70\,\micron,
slightly higher than those quoted by \citet{engelbracht08} and \citet{herrera12},
respectively.
Our photometry agrees very well with these previous estimates.

\izw\ is clearly detected at 70 (MIPS, PACS) and 100\,\micron\ (PACS). 
At 160\,\micron\ with PACS, the detection is marginal, $\sim2.4\sigma$ 
(see Fig. \ref{fig:pacsizw}).
Our photometry is consistent with that reported by \citet{remy13}
for 70 and 100\,\micron.
Using shallower images, both \citet{remy13} and \citet{herrera12} 
report non-detections at 160\,\micron:
the former give a 5$\sigma$ upper limit of 11\,mJy, and the latter
a 3$\sigma$ limit of 27.2\,mJy, both consistent with our measurement
based on deeper data.

Basic data for \izw\ are reported in Table \ref{tab:basicizw}, and
the compiled SED data are given in Table \ref{tab:sedizw}.
At the assumed distance of 18.2\,Mpc, 1\arcsec\,=\,88\,pc.

\subsection{Fitting the SED of I\,Zw\,18}
\label{sec:sedizw}

As for \sbs, before fitting the SED, it was necessary to correct
the photometry from 1$-$10\,\micron\ for nebular emission.
Based on \izw's SFR\,=\,0.17\,\msunyr\ \citep[][scaled to D\,=\,18.2\,Mpc]{hunt05b},
this turns out to be an important correction:
$\sim$50\% of the 3.6\,\micron\ and 4.5\,\micron\ emission is from nebular continuum $+$ \bra.
\izw\ is extended at short IRS wavelengths \citep[see][]{wu07}, 
and judging from the IRAC global photometry,
the 3\farcs6 aperture for the IRS short-low spectral range
misses 50\% or more of the emission.
Hence, the free-free subtraction resulted in unrealistically low fluxes,
so we did not consider in the SED fitting
the short-wavelength IRS fluxes $\la10$\,\micron.
The 2.6\,mm flux observed by \citet{leroy07} is also contaminated by
free-free emission; extrapolating from the thermal fraction (47\%)
of the 8.4\,GHz radio continuum \citep{hunt05b,cannon05} gives
a free-free component at 115\,GHz of 0.37\,mJy.
\citet{leroy07} estimate that the free-free flux at 115\,GHz is between
0.36 and 0.51\,mJy, which means that from 35\% to 48\% of
the observed flux [(1.06\,$\pm$\,0.35)\,mJy] is bremsstrahlung.

\begin{figure}[ht]
\centerline{
\includegraphics[angle=0,height=\linewidth]{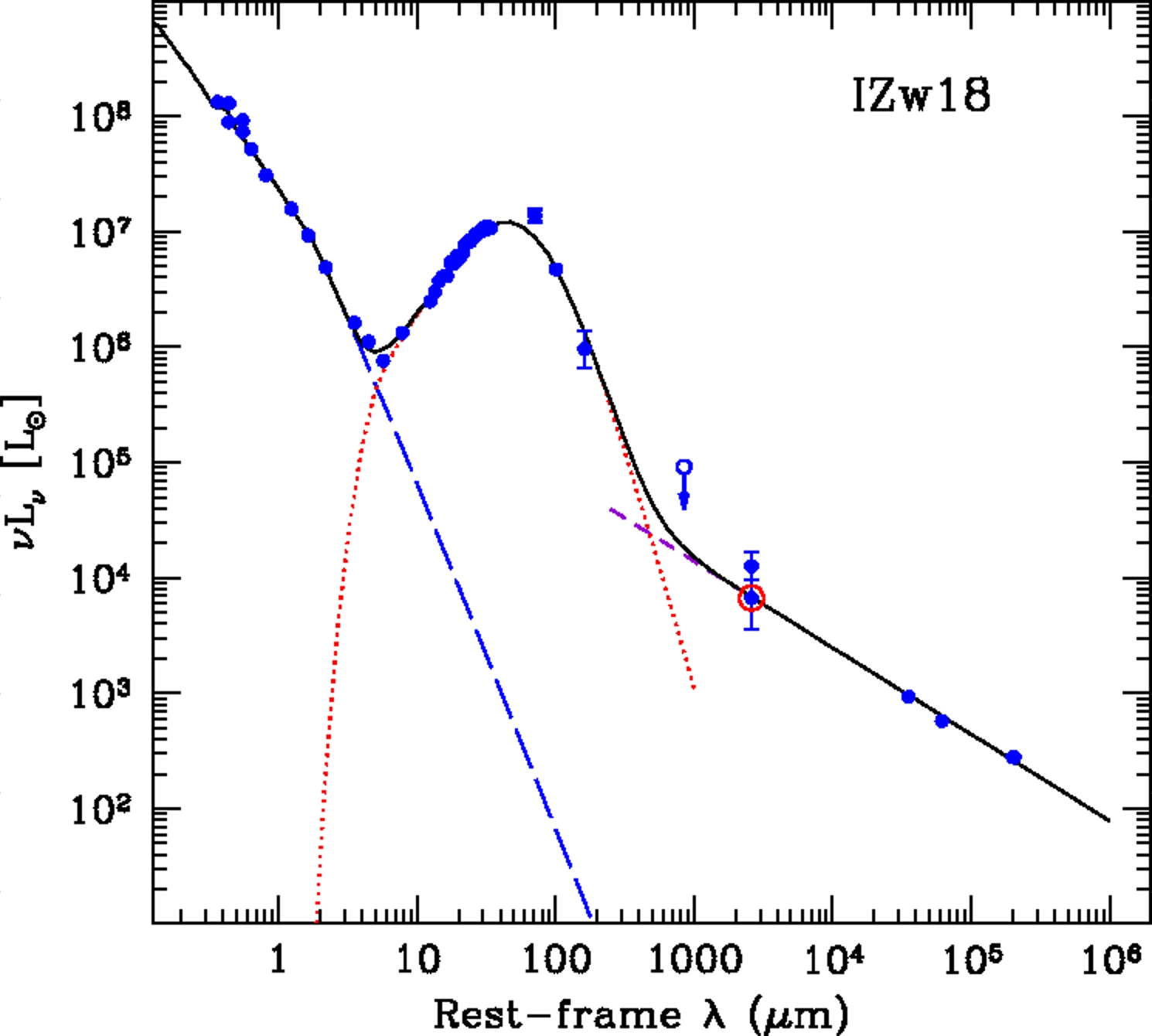}
}
\caption{SED of \izw, together with best-fit \dusty\ model overlaid.
The large open (red) circle shows the dust-only emission taken from \citet{leroy07},
while the small filled (blue) circle above it shows the total 2.6\,mm emission.
The black curve represents the \dusty\ fit, and the dotted (red) 
curve, only the dust; the long-dashed (blue) curve shows the stars.
The error bars are masked by the size of the symbols.
As in Fig. \ref{fig:sedsbs},
\dusty\ models only extend to 1\,mm, so the excess above the dust-only
emission is the extrapolation of the radio with a spectral index of $-0.25$
(see text).
The pure radio component is shown by a short-dashed (purple) line
(longward of 250\,\micron).
}
\label{fig:sedizw}
\end{figure}

The \spit/IRS spectrum (not included in Table \ref{tab:sedizw}) is taken from \citet{wu07},
and it was averaged over 1\,\micron\ bins after subtracting emission lines.
As for \sbs, to fit the SED we imposed only a prior on metallicity.
The result of the SED fitting is shown in Fig. \ref{fig:sedizw}.
The radio emission is approximated by a radio
spectral index of $\alpha\,=\,-0.25$, slightly steeper than pure thermal
emission because of a global synchrotron component \citep{hunt05b}.

%
\begin{center}
\begin{table}
      \caption[]{Basic data for \izw} 
\label{tab:basicizw}
\addtolength{\tabcolsep}{7pt}
{
\tiny
\begin{tabular}{llc}
\hline
Parameter  & \multicolumn{1}{c}{Value$^{\mathrm{a}}$} & Reference$^{\mathrm{b}}$ \\
\hline
$V_{\rm hel}$  & 751 km\,s$^{-1}$           & (1) \\
Distance       & 18.2\,Mpc (1\arcsec\ = 88\,pc)   & (2) \\
\logoh         & 7.17 (NW) $-$ 7.18 (SE)    & (3) \\
M$_{\rm stars}$    & $1.34 \times 10^{6}$\,\msun & (4) \\
SFR  	       & 0.17\,\msunyr                  & (5) \\
$\Sigma_{\rm SFR}^{\mathrm{c}}$ & 0.19\,\msunyr\,kpc$^{-2}$  & This paper \\
M$_{\rm H\,I}$ & $1.0 \times 10^{8}$\,\msun  & (6) \\
$\langle\Sigma_{\rm H\,I}\rangle$ & 64\,\msunpc  & (6) \\
Peak $\Sigma_{\rm H\,I}$ & 100\,\msunpc  & (6) \\
L$_{\rm dust}^{\mathrm{d}}$  & ($2.00\,\pm\,0.07 )\times 10^{7}$\,\lsun  & This paper \\
M$_{\rm dust}$  & ($3.4\,\pm\,1.0 )\times 10^{2}$\,\msun  & This paper \\
\hline
\end{tabular}
}
\vspace{0.5\baselineskip}
\begin{description}
\item
[$^{\mathrm{a}}$]
Luminosity and mass values taken from the literature
have been scaled to the distance of $D$ = 18.2\,Mpc. 
\item
[$^{\mathrm{b}}$] (1) NASA/IPAC Extragalactic Database (NED;
(2) \citet{aloisi07};
(3) \citet{izotov99};
(4) \citet{fumagalli10};
(5) \citet{hunt05b};
(6) \citet{lelli12}.
\item
[$^{\mathrm{c}}$]
Obtained by considering an area of 0.88\,kpc$^2$,
corresponding to a circular aperture 6\arcsec\ 
($\sim$530\,pc) in radius.
\item
[$^{\mathrm{d}}$]
Obtained by integrating the SED model described in Sect. \ref{sec:izw}.
\end{description}
\end{table}
\end{center}

%
\begin{center}
\begin{table}
      \caption[]{Global photometry for \izw} 
\label{tab:sedizw}
\addtolength{\tabcolsep}{3pt}
{
\tiny
\begin{tabular}{lllc}
\hline
& & \multicolumn{1}{c}{1$\sigma$}\\
\multicolumn{1}{c}{Wavelength}  & \multicolumn{1}{c}{Flux$^{\mathrm{a}}$} & 
\multicolumn{1}{c}{Uncertainty} & Reference$^{\mathrm{b}}$ \\
\multicolumn{1}{c}{(\micron)}  & \multicolumn{1}{c}{(mJy)} & 
\multicolumn{1}{c}{(mJy)} \\
\hline
0.366 &   1.56  &  0.223  & (1) \\
0.439 &   1.25  &  0.051  & (2) \\
0.440 &   1.82  &  0.250  & (1) \\
0.553 &   1.63  &  0.234  & (1) \\
0.555 &   1.31  &  0.055  & (3) \\
0.641 &   1.06  &  0.063  & (2) \\
0.815 &   0.80  &  0.035  & (3) \\ 
1.25 &    0.63$^\mathrm{c}$  &  0.028  &  (3) \\
1.65 &    0.49$^\mathrm{c}$ &  0.025  &  (3) \\
2.20 &    0.35$^\mathrm{c}$  &  0.019  &  (3) \\
3.55   & 0.19$^\mathrm{c}$ &  0.012 &  (4) \\
4.49   & 0.16$^\mathrm{c}$ &   0.012 &  (4) \\
5.73   & 0.14$^\mathrm{c}$ &   0.038 &  (4)  \\
7.87   & 0.34$^\mathrm{c}$  &  0.037    & (4) \\ 
23.7  &   6.2  &  0.31 & (4) \\
23.7  &   6.3  &  0.68 & This paper (MIPS$^{\mathrm d}$) \\
71.4  &  31.2   & 4.4   & This paper (MIPS$^{\mathrm d}$)  \\ 
71.4  &  33.6   & 1.70   & (5) \\ 
71.1   &   32.0   &  4.3    & This paper (PACS$^{\mathrm d}$) \\ 
101.2  &   15.3   &  0.9    & This paper (PACS$^{\mathrm d}$) \\ 
162.7  &   5.0   &  2.2    & This paper (PACS$^{\mathrm d}$)\\ 
849.2  &  2.5$^{\mathrm{e}}$    &  ...   & (6) \\ 
2607   &  1.06$^{\mathrm{f}}$   &  0.35  & (7) \\ 
2607   &  0.56$^{\mathrm{g}}$   &  0.49  & (7) \\ 
35989  &  1.06   & 0.08 &  (8) \\
61686  &  1.14   & 0.14 &  (8) \\
199862 &  1.79   & 0.18 &  (8) \\
205337 &  1.83   & 0.33 &  (8) \\
\hline
\end{tabular}
}
\vspace{0.5\baselineskip}
\begin{description}
\item
[$^{\mathrm{a}}$] 
Fluxes given here are corrected for Galactic extinction, \av\,=\,0.106\,mag.
\item
[$^{\mathrm{b}}$] 
(1) \citet{papaderos02};
(2) \citet{gildepaz03};
(3) \citet{hunt03};
(4) \citet{engelbracht08};
(5) \citet{herrera12};
(6) \citet{hunt05a};
(7) \citet{leroy07};
(8) \citet{hunt05b}.
\item
[$^{\mathrm{c}}$]
Corrected for nebular continuum as described in the text.
\item
[$^{\mathrm{d}}$] 
Photometry in an aperture of 13\,\arcsec\ radius.
\item
[$^{\mathrm{e}}$] 
3$\sigma$ upper limit.
\item
[$^{\mathrm{f}}$] 
Total observed emission.
\item
[$^{\mathrm{g}}$] 
Dust-only emission as calculated by \citet{leroy07},
see Sect. \ref{sec:sedizw}. 
\end{description}
\end{table}
\end{center}

Like the SED of \sbs, \izw's SED is warm;
it peaks between 50 and 60\,\micron.
The \dusty\ fit fails to fit the estimated dust-only emission at
2.6\,mm, and the prediction of the fit is such that dust would
be impossible to detect at this relatively long wavelength\footnote{The
gradient in the radio spectral index makes the
thermal/non-thermal separation in \izw\ problematic \citep{hunt05b}.}. 
The temperature at the inner radius of the shell is 
850$^{+20}_{-147}$\,K,
warmer than in \sbs; this temperature is the maximum probed by our
\dusty\ template library, so perhaps it is even a lower limit.
The shell is about 5 times thinner than for \sbs, 
$Y\,\equiv\,R_{\rm out}/R_{\rm in}$ 
132$^{+11}_{-18}$ vs. $Y\sim600$.
Unlike \sbs, there is no signature of silicate absorption in the SED
of \izw; in fact the best-fitting dust is that produced by primordial Population III
supernovae \citep{bianchi07} rather than the SMC dust that gave the best
fit for \sbs.
It was impossible to fit the SED of \izw\ with either the Milky Way dust
or the SMC dust as given by \citet{weingartner01}; this could be because of
an intrinsic low silicate abundance in \izw\ since the SNe dust 
used in our models \citep{bianchi07} contains no silicate grains.

The best-fit stellar age is 18.3$^{+1.7}_{-2.5}$\,Myr, $\approx$2\% solar metallicity,
and with no measurable extinction.
This age is compatible with the mix of ages obtained from fitting
the global optical$+$NIR photometry \citep{hunt03}.
The implied stellar mass given by the $V$-band luminosity of $6.56\times10^7$\,\lsun,
and $M/L_V$\,=\,0.028 \citep[for a stellar population 18\,Myr old,][]{bc03} is
$1.82\times10^6$\,\msun, roughly consistent with the stellar mass of
$1.34\times10^6$\,\msun\ found by \citet{fumagalli10} using broad-band colors.
The $V$-band dust optical depth \tauvd\ given by the \dusty\ fit is 72$^{+8}_{-1}$, quite high,
typical of thick clouds.
However, the relative normalizations of the two \dusty\ components
(see Eq. \ref{eqn:dusty}) implies that
only 3\% of the dust intercepts the stellar emission.
Hence, the dust must be highly clumped.

The total dust mass in \izw\ given by the \dusty\ fit is $(3.4\,\pm\,1.0)\times10^2$\,\msun,
an extremely small value.
However, it is roughly consistent with what would be expected compared with \sbs.
The IR luminosity of \izw\ ($2.0\times10^7$\,\lsun) is about 80 times lower
than that of \sbs\ ($1.6\times10^9$\,\lsun, see Tables \ref{tab:basicsbs} and \ref{tab:basicizw}).
This means for the same dust mass-to-light (M/L) ratio (i.e., for the same mean dust temperature),
\izw\ should have $\sim$1/80 of the dust mass of \sbs, $\sim$475\,\msun.
Since \izw\ has a hotter temperature as inferred from the higher value of \tin\ from the \dusty\ fit
(and as discussed in the following), the dust
mass would be even lower (because of the reduced dust M/L), 
thus consistent with the $\sim$340\,\msun\ of dust we obtain from the \dusty\ fit. 

Again, as a ``sanity'' check, we performed a single-temperature
MBB fit of the SED, as shown in the right panel of Fig. 
\ref{fig:MBB}\footnote{The MBB fit assumes that the dust is optically thin, whereas the
\dusty\ fit would imply that the dust is somewhat optically thick.}.
As for \sbs, fixing $\beta$\,=\,2 and letting $T$ vary between 5\,K and 100\,K
in order to apply MIPS and PACS color corrections, we obtain
a best-fit temperature of \izw\ of $T$\,=\,67\,$\pm$\,2.4\,K, warmer than \sbs\ 
($T$\,=\,59\,$\pm$\,1.0\,K).
With the SNe dust emissivity at 100\,\micron\ (a clear detection),
we would estimate dust mass of $\sim$130\,\msun, a factor of $\sim$3 lower than
the \dusty\ value.
Judging from Fig. \ref{fig:MBB},
unlike in \sbs, the single-$T$ MBB fit for \izw\ is missing a small fraction
of dust; in fact, this fit slightly overshoots the IRS portion of the SED.
Nevertheless, both the MBB and \dusty\ fits may be missing some cool dust
mass because of the lack of constraints at longer wavelengths ($\ga$300\,\micron).

There is some degeneracy in the \izw\ \dusty\ fits, in the sense
that large $Y$ ($R_{\rm out}/R_{\rm in}$) and small \taud\footnote{This is
really \tauvd, as defined in Sect. \ref{sec:sed}.}
give a similar SED to small $Y$ and large \taud\ (which is the best-fit dust). 
Such a degeneracy is only possible in the case of featureless SNe dust, because
any dust with significant silicates would produce a deep absorption
feature at $\sim$10\,\micron\ with large \taud.
The large $Y-$small \taud\ fits of \izw\  give a factor of 3 less dust
($\sim$100\,\msun) than small $Y-$large \taud.
The quality assessment of the small $Y-$large \taud\ fit of \izw\ 
shows that it is statistically superior, so the larger dust mass seems
more probable.
Nevertheless, there is significant uncertainty in the dust mass for \izw. 

Our value of \mdust\ for \izw\ is roughly consistent with the measurement
by \citet{fisher13},
given their range of acceptable values and
the difference in dust emissivity for the Milky
Way used by them \citep[MW,][]{draine07} relative to the best-fit SNe dust.
At 100\,\micron\ $\kappa_\nu$ for the MW is $\sim$40\% lower than for the best-fit SNe dust,
causing their lower limit of 450\,\msun\ to become 326\,\msun, close to
our value of 342\,\msun.

\section{Dust and gas in low-metallicity starbursts}
\label{sec:dustgas}

We have found $\sim4\times10^4$\,\msun\ of dust in \sbs, and
an amount of dust more than 100 times lower in \izw. 
Such a result is unexpected because both BCDs have roughly the same metallicity
and are equally rich in gas.
The ratio of \hi\ to stellar mass in \sbs\ is 76 and 75 for \izw.
This means that the baryonic mass is dominated by atomic hydrogen;
\hi\ comprises $\sim$99\% of the total baryonic mass (without
considering a potential molecular gas component), and 
will potentially provide a vast amount of fuel for future star formation.

\subsection{Dust-to-gas ratios}
\label{sec:dustgasratios}

The dust-to-gas mass ratio, \dgr, is a measure of the metals
bound up in dust grains. 
\dgr\ should follow the balance of dust formation processes
and grain destruction from SNe explosions and other violent events in the ISM.
If the interstellar abundances of heavy elements were proportional
to the gas-phase oxygen abundance O/H, we would expect a trend of
\dgr\ with \logoh, under the assumption that the fraction of
metals in dust does not vary from galaxy to galaxy.
Such a correlation has been shown to hold in local galaxies
\citep[e.g.,][]{draine07sings,galametz11,leroy11} and at high redshift
\citep[e.g.,][]{santini10,magdis11}.

\begin{figure}[ht]
\centerline{
\includegraphics[angle=0,width=\linewidth]{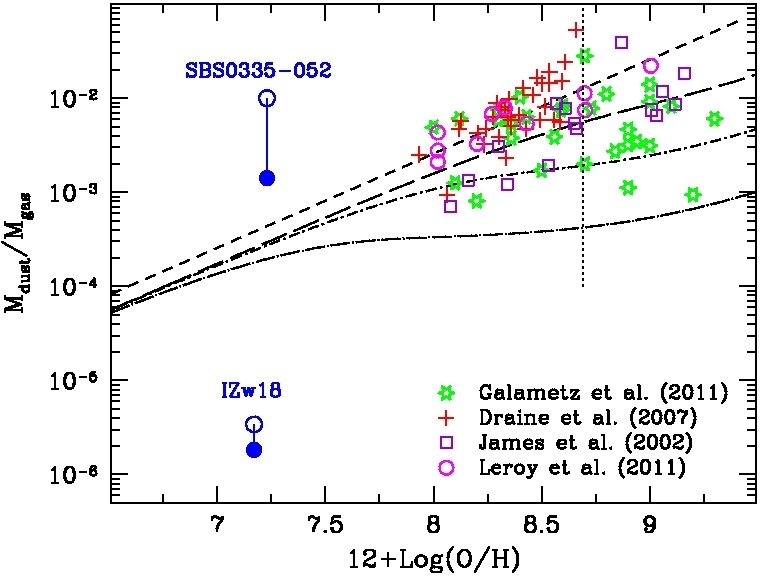}
}
\caption{Dust-to-gas mass ratios \dgr\ as a function of metallicity.
Open circles show \dgr\ inferred from the \hi\ gas mass of \sbs\ and \izw; 
the filled (blue) circles give the total gas mass
after adding the putative molecular component discussed in Sect. \ref{sec:dustgasratios}.
Galaxies from \citet{galametz11} are shown as (green) open stars,
those from \citet{draine07sings} as (red) $+$,
from \citet{james02} as (purple) squares,
and from \citet{leroy11} as (magenta) open circles.
The short dashed line is the linear prediction by \citet{draine07sings},
and the three dot-dashed curves below show the models by \citet{hirashita02}.
The vertical dotted line illustrates solar metallicity \logoh\,=\,8.69 \citep{asplund09}.
}
\label{fig:d2g}
\end{figure}

While measurements of dust-to-gas ratios at low metallicity
are becoming more frequent \citep[e.g.,][]{chen13}, up to now
there have been no reliable estimates of dust masses at \logoh$\la$8
\citep{draine07sings,galametz11}. 
Here we add \sbs\ and \izw\ to the available datasets, and test whether the correlation
between \dgr\ and O/H extends to these low abundances.
The main problem with such a test is that the total gas mass estimate is missing 
the molecular component;
this point will be discussed further in Sect. \ref{sec:alphaco}.
If we take the total \hi\ mass in \sbs\ from \citet{ekta09},
integrated over a region $\ga$2.5\,kpc$^2$ (several $4\times3$\,arcsec$^2$ beams, see Table \ref{tab:basicsbs}),
we obtain \dgr\,=\,$8.9\times10^{-5}$, which would be well below the linear
trend of \dgr\ with O/H.
However, as noticed by \citet{draine07sings}, it is important
to consider the gas mass in the same region as the dust;
otherwise \dgr\ will be unfairly underestimated.
Therefore, to calculate the gas mass we relied on column densities
rather than the total \hi\ masses given in Tables \ref{tab:basicsbs} and \ref{tab:basicizw}.
The dust in \sbs\ subtends an area of $\sim$1\,arcsec$^{2}$,
and the \hi\ column density (measured in UV absorption)
in a $2\times2$\,arcsec$^{2}$ aperture is $\sim$56\,\msunpc\ \citep{thuan97a}.
Hence, the total \hi\ mass in the region where dust coexists is 
$3.85\times10^6$\,\msun, and \dgr\,=\,0.0099.
This gives the same \dgr\ as dividing
the dust column density (0.55\,\msunpc) by the \hi\ column (56\,\msunpc).
Applying the same reasoning to \izw, we find \dgr\,=\,$3.4\times10^{-6}$,
assuming that both dust and gas subtend a circular region 8\arcsec\ (700\,pc)
in radius.
We would obtain the same value if we divided the total dust mass ($3.4\times10^2$\,\msun)
by the total \hi\ mass ($1\times10^8$\,\msun),
implying mean dust and \hi\ column densities in 
\izw\ of $2.2\times10^{-4}$\,\msunpc\ and 64\,\msunpc, respectively.

Figure \ref{fig:d2g} plots dust-to-gas mass ratios for four samples 
of galaxies taken
from the literature, together with our new estimates for \sbs\ and \izw.
The linear trend by \citet{draine07sings} shown as a short-dashed line fits the 
metal-rich data quite well.
However, 
with \hi\ alone (shown as blue open circles in Fig. \ref{fig:d2g}), 
\sbs\ exceeds the trend by a factor of 30 or more,
but \izw\ lies almost a factor of 100 below the trend.
If our estimate of dust mass in \izw\ is correct,
at low metallicity the trend of abundance and dust-to-gas ratios changes
drastically, and in different ways according to some other parameter,
for example SFR surface density or column densities of dust and ionized gas. 

\begin{figure}[ht]
\centerline{
\includegraphics[angle=0,width=\linewidth]{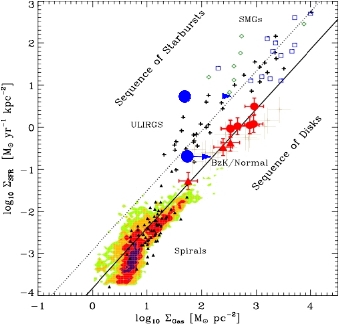}
}
\caption{\sbs\ (upper filled blue circle) and \izw\ (lower filled blue circle)
on the gas scaling relations given by \citet{daddi10}.
Only \hi\ mass density is considered for the BCDs studied here;
the arrows correspond to the missing gas mass surface density that would
transport \sbs\ to the ``starburst sequence'' and \izw\ to the ``sequence of
disks''.
SFR surface density is plotted as a function of the gas (atomic and molecular) surface
density for $z\sim0.5$ disks (shown as red filled circles and triangles;
brown crosses indicate $z=1-2.3$ normal galaxies, and
empty squares correspond to sub-millimeter galaxies.
The shaded regions and filled triangles are local spiral galaxies, and 
small crosses are starbursts and ultra-luminous galaxies.
References for these data are given in \citet{daddi10}.
}
\label{fig:daddi}
\end{figure}

Our ALMA observations did not detect CO(3-2) in \sbs, but either the expected
linearity in \dgr\ is incorrect, or there is a large fraction of missing
molecular gas.
We can use gas scaling relations
to estimate the total gas content in \sbs\ and check whether we are also
missing a molecular component in \izw.
If the SF in \sbs\ occurs in the roughly same area as the ALMA emission
($\sim$1\,arcsec$^2$), we would find \sigmasfr$\sim$18\,\msun\,kpc$^{-2}$.
However, if we conservatively consider a larger area of diameter 550\,pc
($\sim$2\farcs1), comparable to the region emitting 
faint \pa\ \citep{reines08}, we obtain \sigmasfr$\sim$5\,\msun\,kpc$^{-2}$.
The SF in \izw\ takes place in a region of diameter $\sim$1\,kpc
\citep{cannon02,cannon05}, and we
find \sigmasfr$\sim$0.15\,\msun\,kpc$^{-2}$. 
These spatial scales are sufficiently large that the gas scaling relations are
not expected to have broken down \citep[e.g.,][]{verley10,onodera10}.
Fig. \ref{fig:daddi} shows SFR surface density plotted against gas surface
density; the figure is taken from \citet{daddi10}.
\sbs\ (the upper point) and \izw\ (the lower one) correspond to filled blue circles, 
and the arrows indicate
how much gas would be needed to bring \sbs\ to the starburst sequence
and \izw\ to the sequence of disks. 
At \sbs's SFR surface density, 
the starburst sequence of the SFR-gas scaling would require
a total gas surface density of $\sim$400\,\msunpc.
\izw\ already has too much \hi\ to fall on the starburst sequence;
if, on the other hand, it were to fall on the disk sequence,
it would need a total gas surface density of $\sim$158\,\msunpc.
The missing gas mass, which we take as the missing molecular component,
is $\sim$342\,\msunpc\ for \sbs, and $\sim$94\,\msunpc\ for \izw,
assuming for the latter the mean \hi\ column density of 64\,\msunpc\
(see Tables \ref{tab:basicsbs} and \ref{tab:basicizw}). 
However, if \izw\ falls on the starburst sequence, similarly to what we
hypothesize for \sbs, it would be missing no molecular mass;
indeed, additional gas would move \izw\ farther from what we might
expect.
In any case,
these estimates of missing molecular gas for both galaxies are highly uncertain,
and should be taken as a basis for discussion, rather than conclusive
determinations.

Adding the (putative) missing molecular gas gives a total gas mass
that corresponds to the filled circles in 
the dust-to-gas mass ratios plotted in Fig. \ref{fig:d2g}.
The missing molecular component would be 
$\sim$1.5 times the \hi\ mass in \izw\ but
a factor of 6 times the \hi\ mass in \sbs\ (missing 342\,\msunpc\ vs. 56\,\msunpc\ in \hi).
\sbs\ would be moved closer to, and possibly consistently with, the linear trend
of \dgr\ with metallicity, while \izw\ would be moved farther away. 

We therefore tentatively conclude that there is a large component of missing
molecular gas in \sbs; in \izw\ such a conclusion seems more remote, though possible.
\citet{krumholz12} argues that at low metallicities
(perhaps lower than in these galaxies), star formation will occur in 
the cold atomic phase of the ISM rather than necessarily in a molecular phase.
This could be the case in \sbs\ and \izw\, 
since in both galaxies
star formation is taking place in a very \hi-rich medium.
Nevertheless, the total gas fractions could be even higher in \sbs;
if \sbs\ is to fall on the linear trend of \dgr\ with O/H,
it would be missing a yet unmeasured gas component.
We discuss such a gas component in the next section.

\subsection{Constraints on molecular gas conversion factors in \sbs\ and \izw}
\label{sec:alphaco}

Previous attempts to detect \coone\ and \cotwo\footnote{Hereafter
referred to without the isotopic prefix.} in \sbs\ were carried out
by \citet{dale01} at Owens Valley. After 20.6\,hr of on-source integration
time at 3mm and 6.6\,hr of good integration time at 1mm, they achieved
3$\sigma$ upper limits of 1.2\,Jy\kms\ and 3.0\,Jy\kms, respectively.
Converting these to brightness temperatures gives 3$\sigma$ flux upper limits
of 5.6\,\kkms\ for CO(1-0) and 12.2\,\kkms\ for CO(2-1).
We can use these limits to constrain values for
\aco, 
the factor to convert observed CO intensity $I_{\rm CO}$ to \htwo\ column density.
In order to account for the ``missing'' molecular gas density of
342\,\msunpc\ in \sbs, \aco\ would be $\ga$61\,\msunpc\,(\kkms)$^{-1}$
for CO(1-0), 10-12 times higher than typical values for
the Milky Way \citep[e.g.,][]{bolatto13}, but lower than would be expected
($\sim$30 times higher) for a linear trend of \aco\ with metallicity \citep[e.g.,][]{draine07sings}.
Assuming a CO(2-1)/CO(1-0) ratio of 0.7 \citep{leroy11}, we would obtain
\aco$\ga$40\,\msunpc\,(\kkms)$^{-1}$ for CO(2-1).

More stringent lower limits can be achieved if
we use to estimate \aco\ the 3$\sigma$ CO(3-2) upper limit
for \sbs\ (1.58\,\kkms) given in Sect. \ref{sec:data}. 
Because \sigmahtwo\,=\,\aco\ $I_{\rm CO(1-0)}$, we need first to
convert the CO(3-2) limit to one on CO(1-0).
For a sample of dwarf starbursts,
\citet{meier01} found an error-weighted mean CO(3-2)/CO(1-0) flux ratio
of 0.6$\pm$0.06, roughly consistent with the 
median value of 0.7 obtained by \citet{mao10}.
Converting CO(3-2) to CO(1-0) using the range in these ratios, to accommodate the ``missing'' 342\,\msunpc\ 
gas density in \sbs, \aco$\ga$130-152\,\msunpc\,(\kkms)$^{-1}$.

The same exercise can be performed for \izw, using the 3$\sigma$
CO(1-0) limit of 0.75\,\kkms\ given by \citet{leroy07}\footnote{They
find a 4$\sigma$ upper limit of 1\,\kkms\ which we convert to 3$\sigma$
for consistency with our \sbs\ calculation.}. 
From Fig. \ref{fig:daddi},
we estimate that \izw\ is missing $\sim$94\,\msunpc\ of gas density;
hence \aco$\ga$125\,\msunpc\,(\kkms)$^{-1}$.
The two lower limits for \sbs\ and \izw\ are similar,
but this is merely fortuitous, since it depends on the sensitivity of
the CO observations, the missing gas,
and the flux ratio of CO(3-2)/CO(1-0) in the case of \sbs. 
In the case of \sbs, the most conservative hypothesis
would be that more gas is needed to bring it onto the starburst sequence of gas scaling
relations (rather than the quiescent sequence).
However, \izw\ is already on the starburst sequence with
only \hi; putative missing gas would be needed to move this galaxy onto
the more quiescent sequence of disks.
If no molecular gas component were needed for \izw\ in the gas scaling
relations (i.e., if it really should be on the starburst sequence), 
\aco\ for \izw\ would be undefined.

\begin{figure}[ht]
\centerline{
\includegraphics[angle=0,width=\linewidth]{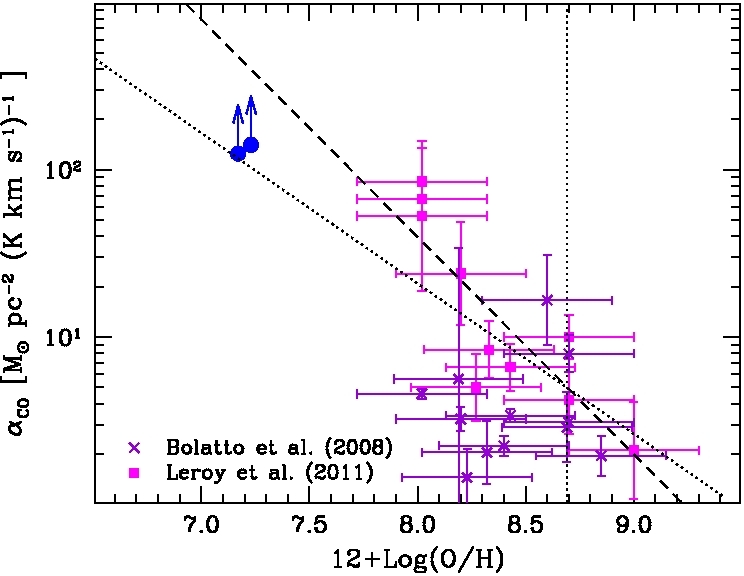}
}
\caption{\aco\ vs. \logoh\ for nearby galaxies.
The lower limits for \sbs\ and \izw\ are shown as filled blue
circles; data from \citet{leroy11} and \citet{bolatto08} are shown
as filled squares and $\times$, respectively.
The steeper regression, indicated by a dashed line,
has a slope of $-1.3$ \citep{genzel12}
and the shallower one, shown by a dotted line, a slope of $-0.9$ \citep{sargent13}.
}
\label{fig:alphaco}
\end{figure}

Figure \ref{fig:alphaco} shows the \aco\ lower limits for \sbs\ and \izw,
together with measured \aco\ conversion factors for two samples of nearby
galaxies \citep{bolatto08,leroy11}.
The steeper of the two regressions (slope $-1.3$) is given by \citet{genzel12},
and the shallower one (slope $-0.9$) by \citet{sargent13}.
Our lower limits on \aco\ for \sbs\ and \izw\ lie slightly above the shallower
regression, which would imply that the slope is closer to linear, or even
super-linear as found by \citet{genzel12}.
To better constrain the low-metallicity end of the \aco\ conversion factor,
deeper observations are needed.
Even non-detections, but at a fainter intensity level, will
help understand the fate of molecular gas and its tracers at low metallicity.

\section{Summary and conclusions}
\label{sec:conclusions}

We have presented our first ALMA results for \sbs, one of the most metal-poor
star-forming dwarf galaxies in the Local Universe.
The ALMA Band 7 continuum observations at 870\,\micron\ allow us to reach an unprecedented spatial resolution of 90\,pc, 
even with the Cycle 0 capabilities.
Extrapolating from high-resolution high-frequency radio observations, we find
that 85\% of the Band 7 flux in \sbs\ can be attributed to free-free emission.
The remainder is attributed to dust, and we use this new constraint
to estimate the dust mass in \sbs\ by fitting its SED with a \dusty\ model.
The ratio of the resulting dust mass of $3.8\times10^4$\,\msun\  in \sbs\ 
and its stellar mass falls within the range of normal galaxies.
However, comparing the dust with \hi\ mass
and surface density, we find that \sbs\ far exceeds the linear trend
of dust-to-gas mass ratio \dgr\ with oxygen abundance.

We have compared the results for \sbs\ to another dwarf galaxy at similarly low
metallicity, \izw.
With data from the literature, and \hers\ archival observations,
we have estimated the dust mass in \izw\ by fitting
a \dusty\ model, and find $3.4\times10^2$\,\msun\ of dust.
Compared with its \hi\ mass and column density, unlike \sbs,
\izw\ falls far below the expected scatter of the linear trend of \dgr\ with O/H.
Finally, we have used the CO non-detections for both galaxies to constrain
the integrated CO(1-0) intensity to \htwo\ surface density conversion factor, \aco, and find that 
the lower limits are
consistent with a linear, or perhaps even super-linear trend with metallicity.

Perhaps the most significant of our findings is that metallicity does not
uniquely define star-formation processes.
Considering only \hi\ gas,
\izw\ falls roughly on the starburst sequence of gas scaling relations, but
\sbs\ is a clear outlier; a gas surface density 6 times higher than that from
\hi\ alone would be required to bring it onto
the sequence of starbursts (see Fig. \ref{fig:daddi}).

Although specific SFRs (SFR/\mstar) are similar for both galaxies ($\sim 10^{-7}$\,yr$^{-1}$), the
surface densities of SF are very different.
SF in \sbs\ occurs mainly in two luminous compact super-star clusters, while
that in \izw\ originates in more diffuse ones;
the SFR surface densities differ by more than an order of magnitude.
This is probably the reason for the extreme short-wavelength peak in \sbs's SED
($\sim20-30$\,\micron) and implies that physical conditions are extreme in this galaxy.
Considering also the large volume densities of ionized gas
implied by its self-absorbed thermal radio spectrum \citep{hunt04,johnson09},
the star formation activity in \sbs\ must be occurring in very warm, dense
regions.
The physical conditions in \izw\ are less extreme;
the radio spectrum is not self-absorbed, and electron densities are lower.

Ultimately, 
the influence of metallicity may not be as clear cut as commonly thought.
At some level, self-shielding and the survival of molecules may depend
just as much on gas and dust column density as on metallicity.
The effects of low metallicity may at least be partially compensated for
by large column densities in the ISM.
More observations of extreme metal-poor galaxies are needed to judge
whether this is generally true, and to better understand the degree
to which such galaxies can serve as proxies for galaxy formation at high 
redshift.

\begin{acknowledgements}
We appreciate the prompt review and careful
comments of the referee which improved the clarity of the paper.
We warmly thank Bruce Draine for constructive comments and suggestions,
and Emanuele Daddi for interesting discussions. 
This paper makes use of the following ALMA data: ADS/JAO.ALMA\#2011.0.00039.S (PI Hunt). 
ALMA is a partnership of ESO (representing its member states), NSF (USA) and NINS (Japan), together with NRC (Canada) and NSC and ASIAA (Taiwan), 
in cooperation with the Republic of Chile. 
The Joint ALMA Observatory is operated by ESO, AUI/NRAO and NAOJ.
We are grateful to the Italian Alma Regional Center for assistance with
data handling and analysis, and to the
International Space Science Institute for hospitality during the conception
of the paper.
LKH acknowledges support from PRIN-INAF 2012/13.
We made use of the NASA/IPAC Extragalactic Database (NED).
\end{acknowledgements}

\end{document}